\documentclass[journal]{IEEEtran}

\usepackage{graphicx}
\usepackage{float}
\usepackage{subfigure}
\usepackage{amssymb}
\usepackage{amsmath}
\usepackage{braket}
\usepackage{comment}

\begin{document}

\title{Challenge-Response Quantum Reinforcement Learning with Application to Quantum-Assisted Authentication}


\author{\IEEEauthorblockN{Jawaher Kaldari\IEEEauthorrefmark{1}, Saif Al-Kuwari\IEEEauthorrefmark{1}}

\IEEEauthorblockA{\IEEEauthorrefmark{1} Qatar Center for Quantum Computing, College of Science and Engineering, Hamad Bin Khalifa University, Doha, Qatar }

Emails: \{jaka51804, smalkuwari\}@hbku.edu.qa

}

\maketitle

\begin{abstract}
Quantum reinforcement learning (QRL) has emerged as a promising research direction that integrates quantum information processing into reinforcement learning frameworks. While many existing QRL studies apply quantum agents to classical environments, it has been realized that the potential advantages of QRL are most naturally explored in environments that exhibit intrinsically quantum characteristics, where the agent's observations and interactions arise from quantum processes. In this work, we propose a quantum reinforcement learning environment formulated as a challenge-response task with hidden information. In the proposed environment, Alice encodes a classical bit into the parameters of a quantum circuit, while Bob, with a trained reinforcement learning agent, interacts with a limited number of quantum state copies to infer the hidden bit. The agent must select measurement strategies and decide when to terminate the interaction under explicit resource constraints. To study the solvability of the proposed environment, we consider three agents: a purely classical agent, a lightweight hybrid agent and a deep hybrid agent. Through experiments, we analyze the trade-off between inference accuracy and quantum resource consumption under varying interaction penalties. Our results show that the lightweight hybrid agent achieves reliable inference using as few as two quantum state copies, outperforming both the classical baseline and the deep hybrid agent across both high and low-penalty regimes. We further evaluate robustness under realistic quantum noise models and discuss the relevance of the proposed environment for security-oriented applications, including quantum-assisted authentication. 
\end{abstract}


\IEEEpeerreviewmaketitle

\section{Introduction}

\IEEEPARstart{R}{ecent} major breakthroughs in artificial intelligence (AI) have been predominantly driven by advances in both deep neural networks and reinforcement learning (RL). Many AI applications rely on deep neural networks, which have transformed the way machines learn and process information, enabling major advances in areas such as image and  speech recognition \cite{krizhevsky2012imagenet,deng2013recent}, fraud detection \cite{zakaria2025detecting}, recommendation systems \cite{rodrigues2026ai}, and drug discovery \cite{wang2025drug}. Although deep neural networks excel at extracting patterns from large, static datasets, they are trained on fixed data and lack the ability to learn and explore through direct interaction with an environment. Therefore, they struggle in settings that require exploration, adaptation to uncertainty, and sequential decision-making. 

RL addresses these limitations by allowing agents to interact with their environment, learn from trial and error, and improve their behavior over time based on feedback in the form of rewards. For example, training a robot using only a deep neural network would require collecting a huge amount of data for every possible situation that a robot might encounter. On the other hand, an RL agent can autonomously explore its environment and refine its policy through experience, making RL especially suitable for dynamic scenarios. RL extends beyond robotics, excelling in areas such as autonomous systems \cite{devi2025reinforcement}, resource allocations \cite{hady2025multi}, and cybersecurity \cite{finistrella2025multi}.

In parallel with these revolutionary advancements, quantum computing has emerged as a new paradigm that harnesses the principles of quantum mechanics to solve certain classes of problems that are believed to be intractable for even the most powerful classical supercomputers. However, large-scale fault-tolerant quantum computers remain a work in progress. Today, we are in the noisy intermediate-scale quantum (NISQ) era, where devices are limited by low qubit counts and short coherence times. Despite these constraints, quantum machine learning (QML) has emerged as one of the most promising research directions to explore the potential of near-term quantum devices.

Within the broader landscape of QML, quantum reinforcement learning (QRL), where RL and quantum computing intersect, has attracted increasing attention in recent years. By leveraging quantum features, such as superposition and entanglement, QRL aims to enhance learning efficiency, exploration strategies, and decision-making in complex environments \cite{chen2026quantum}. Broadly, research in QRL can be categorized into two directions. One prominent line of work focuses on representing the policy itself using parameterized quantum circuits (PQCs), in which quantum circuits act as the policy function governing decision-making. An equally important direction considers RL agents (classical or quantum) interacting with quantum environments, where RL is used to optimize quantum systems, including applications in quantum control \cite{li2025robust}, quantum error correction \cite{nautrup2019optimizing}, and quantum architecture design \cite{ye2021quantum}.

Despite the growing interest in QRL, it remains unclear whether QRL can consistently outperform classical RL beyond highly controlled or artificial settings \cite{kruse2025benchmarking}. In addition, the relative promise of the QRL is still an open question. This has motivated efforts to design systematic evaluation methodologies and benchmarking environments that enable statistically meaningful comparisons of learning performance between classical and quantum RL algorithms \cite{meyer2025benchmarking}. 

In response to these challenges, several works have investigated QRL performance in a different RL environments. For example, QRL has been applied to solve maze-based navigation environments, which are often used as simplified testbeds for analyzing learning behavior \cite{dalla2022quantum, sefrin2022quantum}. In \cite{freinberger2024quantum}, hybrid quantum agents were applied to high-dimensional RL environments such as Atari games, where the proposed hybrid model was capable of solving Pong and achieving performance comparable to a classical baseline in Breakout. QRL has also been explored as a modeling framework in cognitive science, such as in \cite{li2020quantum}, where QRL-based models demonstrated competitive performance compared to classical RL models in explaining behavioral data. Quantum-inspired and quantum-enhanced reinforcement learning approaches have also been explored in healthcare decision-making, where quantum optimization techniques are used to improve policy learning for personalized treatment planning \cite{saranya2025enhanced}. In addition, quantum-inspired reinforcement learning has been investigated in classical environments modeling complex real-world decision problems, such as secure and sustainable supply chain systems \cite{dastagir2025quantum}.

From an RL perspective, environments that are inherently quantum in nature have been identified as natural candidates for quantum agents, as the observations available to the agent arise from measurements on a quantum system and may therefore retain intrinsically quantum characteristics \cite{broughton2020tensorflow}. Fully QRL environments have been introduced in \cite{wu2025quantum}, where both the environment and the interaction process are quantum mechanical in nature, providing a theoretical foundation for agent learning in genuinely quantum settings. Recently, quantum agents have been applied to quantum architecture and circuit search problems, where QRL is used to sequentially select quantum gates and operations to construct high performance target quantum circuits \cite{chen2023quantum}. 

While QRL has been widely applied to quantum systems for control and optimization, to the best of our knowledge, existing quantum RL environments do not explicitly formulate the interaction as a challenge-response mechanism with hidden information as the central objective.

\subsection{Contributions}

Motivated by these considerations, this work introduces a quantum environment and investigates its solvability using agents with varying degrees of quantum involvement. The contributions of this work are summarized as follows:
\begin{itemize}
    \item We propose a novel quantum RL environment that generates a challenge-response task, in which information is embedded in the circuit parameters and remains hidden from the agent. The agent interacts with the environment through a sequence of actions and observations, aiming to infer and extract the encoded information within a limited number of interactions.
    \item We demonstrate the solvability of the proposed environment using RL by considering different agent types (one classical and two hybrid).
    \item We show that the considered agents maintain strong performance in the presence of noisy quantum channels, highlighting the robustness of the approach under realistic conditions.
    \item We demonstrate the relevance of the proposed environment by proposing a security-oriented application, namely quantum-assisted authentication.
\end{itemize}

By formulating the problem as an RL challenge, the proposed environment provides a unified and controlled setting to evaluate RL behavior in different agent realizations.


\subsection{Organization}
The remainder of this paper is organized as follows. Section~\ref{sec:prelimi} provides background on reinforcement learning and quantum reinforcement learning. Section~\ref{sec:QCRC} presents the quantum challenge–response circuit underlying the proposed environment, followed by a detailed description of the environment in Section~\ref{sec:environment}. Section~\ref{sec:agents} describes our agents. Section~\ref{sec:experimentalsetup} presents the experimental setup, Section~\ref{sec:evaluation} introduces the evaluation metrics, and the experimental results are present in Section~\ref{sec:results}. Section~\ref{sec:applications} discusses a potential application of the proposed environment, and Section~\ref{sec:conclusion} concludes the paper.

\section{Preliminaries}\label{sec:prelimi}

\subsection{Reinforcement Learning}
RL is a learning approach in which an agent learns to act in an environment through trial-and-error interactions. This interaction is typically modeled using a Markov decision process (MDP), which is defined by:
\begin{itemize}
\item $\mathcal{S}$: the state space, representing all possible situations the agent may encounter.
\item $\mathcal{A}$: the action space, specifying the set of actions available to the agent.
\item $\mathcal{P}(s' \mid s,a)$: the state transition function, which gives the probability of transitioning to state $s'$ after taking action $a$ in state $s$.
\item A reward function: provides scalar feedback to guide the agent’s behavior.
\item $\gamma \in [0,1]$: the discount factor that controls the relative importance of future rewards.
\end{itemize}
At each discrete time step $t$, the agent observes the current state $s_t \in \mathcal{S}$, selects an action $a_t \in \mathcal{A}$, and receives a reward $r_t$ as the environment transitions to a new state $s_{t+1}$. The agent’s objective is to maximize the expected return, defined as the discounted cumulative reward:
\begin{equation}
G_t = \sum_{\tau=0}^{\infty} \gamma^{\tau} r_{t+\tau}.
\end{equation}
An episode corresponds to a finite sequence of agent–environment interactions, starting from an initial state and terminating when a predefined stopping condition is reached, such as task completion or a maximum number of time steps. The agent’s behavior is governed by a policy, denoted by $\pi$, which specifies how actions are selected given the current state. In general, policies may be stochastic, mapping each state to a probability distribution over actions:
\begin{equation}
\pi(a \mid s) = \Pr\left( a_t = a \mid s_t = s \right).
\end{equation}
Following a policy $\pi$, the agent samples an action according to $\pi(a_t \mid s_t)$ at each time step. Learning in RL involves optimizing the policy parameters, often parameterized by neural networks, to maximize the expected return over episodes of interaction with the environment.

The agent’s objective is to learn a policy that maximizes the expected cumulative reward. Therefore, it must be able to evaluate the long-term desirability of states and actions. This evaluation is captured through value functions that estimate the expected return associated with states or state-action pairs under a given policy. Two main types of value functions are  used in RL:

\begin{itemize}
    \item \textit{State-Value Function:} measures how good it is for the agent to be in a particular state when following a policy $\pi$. It is defined as the expected return starting from state $s$ and thereafter following policy $\pi$:
    \begin{align}
        v_\pi(s)
        &= \mathbb{E}_\pi \!\left[ G_t \mid S_t = s \right] \nonumber \\
        &= \mathbb{E}_\pi \!\left[ \sum_{\tau=0}^{\infty} \gamma^{\tau} r_{t+\tau} \mid S_t = s \right].
    \end{align}

    \item \textit{Action-Value Function:} evaluates how good it is to take a specific action $a$ in state $s$ and then follow policy $\pi$ thereafter. It is defined as:
    \begin{align}
        q_\pi(s,a)
        = \mathbb{E}_\pi \!\left[ G_t \mid S_t = s, A_t = a \right].
    \end{align}
\end{itemize}

RL algorithms are commonly categorized into value-based and policy-based methods. Value-based methods focus on learning value functions, such as the action-value function \(q_\pi(s,a)\), and derive a policy implicitly by selecting actions that maximize the expected long-term return of taking specific actions in given states, as in Q-learning methods \cite{ramesh2025comparative}. In contrast, policy-based methods directly optimize the policy itself, without requiring an explicit value function, as in policy gradient methods \cite{castellini2025difference}.

Hybrid approaches, such as actor-critic methods, combine the strengths of both methods by directly optimizing a policy while using a learned value function to guide and stabilize policy updates \cite{cao2025driftshield}.

\subsection{Quantum Reinforcement Learning}

QRL extends classical RL by incorporating quantum information processing into the learning pipeline. In this setting, the agent’s policy is modeled using a parameterized quantum circuit rather than a classical neural network and is represented by a unitary transformation $U(\theta)$ with trainable parameters $\theta$. 

At each time step, a classical observation or state $s$ is mapped to a quantum state through an encoding operation. This encoding step is required whenever the environment provides classical observations, even if the underlying environment is quantum but produces classical measurement outcomes, and it enables the agent to process classical information using a quantum circuit. Common encoding strategies include rotation-based encodings, where components of the observation vector are embedded into the quantum state via single-qubit rotation gates (e.g., $R_x$, $R_y$, or $R_z$). After encoding, the quantum state is followed by the application of the parameterized unitary $U(\theta)$. The quantum state is evolved as:
\begin{equation}
U_{\theta} \ket{\boldsymbol{\psi}_t}
\end{equation}

Measurement of the resulting quantum state produces classical outputs that are used to define the agent’s action selection.

In contrast to settings with classical observations, some QRL environments, such as quantum control systems in \cite{wu2025quantum}, provide quantum states directly, allowing the agent to process them without an explicit classical-to-quantum encoding.

Like classical RL, QRL algorithms can be broadly categorized into value-based and policy-based approaches. Building on the classical structure of Q-learning, the authors in \cite{chen2020variational} formulated the Frozen Lake and cognitive-radio problems as value-based RL tasks and employed parameterized quantum circuits to approximate the action–value function. More complex architectures have been introduced in \cite{lockwood2020reinforcement} and \cite{skolik2022quantum}.

In contrast to value-based QRL algorithms, policy gradient methods have also been explored in the literature, such as in \cite{jerbi2021parametrized}.  In many cases, the gradient estimates obtained in policy gradient methods can exhibit high variance, which may lead to instability during training. To mitigate this issue while preserving unbiasedness, a baseline term is commonly subtracted from the return \cite{chen2024introduction}. The baseline can be chosen as a value function; in this case, the resulting method is commonly referred to as the advantage actor–critic approach \cite{kolle2024quantum}.

In addition, a variety of more complex QRL architectures have been proposed in the literature, including quantum multi-agent reinforcement learning \cite{yun2023quantum}, free-energy reinforcement learning \cite{levit1706free}, quantum variational autoencoder-based reinforcement learning \cite{nagy2025hybrid}, and quantum hierarchical reinforcement learning \cite{zhu2024efficient}.

\section{Quantum Challenge–Response Circuit}\label{sec:QCRC}

\begin{figure*}[t]
    \centering
    \includegraphics[width=0.8\linewidth]{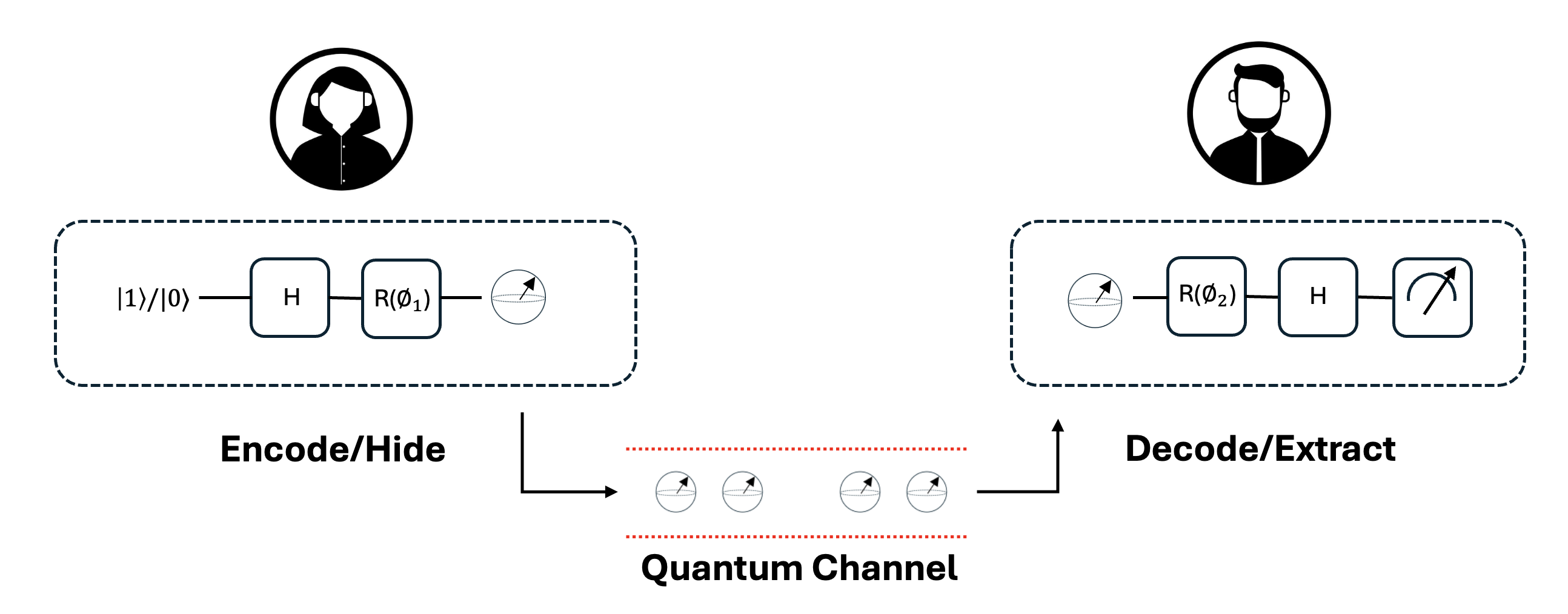}
    \caption{Quantum challenge–response circuit underlying the proposed environment.}
    \label{fig:AliceBobCircuit}
\end{figure*}

Our quantum RL environment is based on a challenge-response mechanism, in which Alice encodes a classical bit into the parameters of a quantum circuit and sends the resulting quantum states to Bob over a quantum channel. Bob, equipped with a trained agent that has learned strategies via RL, manipulates and measures the received qubits to infer and extract the hidden classical bit. Figure \ref{fig:AliceBobCircuit} illustrates the quantum circuit underlying the proposed challenge-response environment.

As illustrated in Fig.~\ref{fig:AliceBobCircuit}, the challenge-response circuit consists of three main steps (to hide and extract information):
\begin{enumerate}
    \item Alice selects a classical bit to encode and prepares the corresponding computational basis state, $|0\rangle$ for bit $0$ and $|1\rangle$ for bit $1$. She then applies a Hadamard gate followed by a parameterized phase rotation gate. Phase rotation is defined by a secret angle $\phi_1$, which is sampled from a predefined range of angles multiple of $\pi$.

    \item Alice sends $N$ identical copies of the resulting quantum state, corresponding to each bit-angle pair, to Bob over the quantum channel.

    \item Bob equipped with a trained RL agent, applies a parameterized phase rotation gate with angle $\phi_2$, followed by a Hadamard gate, and performs measurement. The resulting measurement probabilities depend on the angle difference between Alice’s and Bob’s rotations and are given by
    \begin{equation}\label{probabilities}
    \begin{aligned}
    P(M=1 \mid j=1) = \frac{1 + \cos\!\left(\lvert \phi_1 - \phi_2 \rvert \right)}{2}, \\
    \qquad
    P(M=1 \mid j=0) = \frac{1 - \cos\!\left(\lvert \phi_1 - \phi_2 \rvert \right)}{2}.
    \end{aligned}
    \end{equation}
    Here, $M \in \{0,1\}$ denotes the binary measurement outcome, and $j \in \{0,1\}$ denotes the classical bit encoded by Alice. Through training, Bob learns to adaptively choose $\phi_2$ in order to correctly infer the hidden classical bit using a limited number of quantum state copies. In the ideal case $\phi_1=\phi_2$, the measurement outcome becomes deterministic, yielding outcome “1” with unit probability when $j=1$ and zero probability when $j=0$.

\end{enumerate}

\section{Quantum Environment}\label{sec:environment}

The quantum challenge–response circuit is formulated as an RL environment, in which Bob’s agent is trained to infer the encoded classical bit using a limited number of quantum state copies $N$. By learning an effective policy, the agent enables Alice to transmit only a small number of identical copies per encoded bit; in this work, we demonstrate that reliable inference can be achieved with $N=2$.

In the proposed environment, each hidden bit-angle pair is treated as a single \textit{episode}. At the beginning of the episode, Alice selects a classical bit and a secret angle and prepares $N$ identical copies\footnote{Here, a quantum state copy refers to a single prepared quantum state that can be probed once using a chosen test angle. Repeated measurements (shots) at the same angle are used only to estimate the corresponding measurement probability in equation \ref{observation} and are not counted as additional copies.} of the corresponding quantum state. During the episode, Bob’s agent sequentially selects \emph{test angles} to apply to the received qubits, performs measurements, and records the outcomes. Each interaction consumes one quantum state copy and is treated as a single timestep. This process continues until all $N$ copies are exhausted or the agent decides to terminate the episode. Each interaction (i.e., consumption of one quantum state copy) incurs a fixed penalty $X$, which encourages the agent to minimize the number of qubits used while still achieving reliable inference.
    
At the end of the episode, upon exhausting the available copies or due to early termination, the agent outputs a final guess for the hidden classical bit.  A reward $r$ is assigned based on whether the final guess is correct. The return or cumulative reward for an episode is defined as
\begin{equation}\label{return}
R_{\text{episode}} = r - X \cdot T,
\end{equation}
where $r \in \{+1, -1\}$ denotes the reward corresponding to a correct or incorrect final guess, respectively, $X > 0$ is a fixed probing cost or penalty, and $T$ denotes the number interactions executed prior to termination (i.e., number of quantum state copies used). For example, if the agent takes $T=5$ steps or interactions, then makes a correct guess, but the penalty is $X=0.05$, the total reward would be:
\begin{equation}
    \begin{aligned}
        R_{\text{episode}} &= \underbrace{(-X) + (-X) + \dots + (-X)}_{T=5 \text{ steps}} + \underbrace{(+1.0)}_{\text{correct guess reward}} \\
        R_{\text{episode}} &= (-X \cdot 5) + 1.0 = (-0.25) + 1.0 = 0.75
    \end{aligned}
\end{equation}

Each timestep in the episode has a set of actions and observations available to the agent: 

\begin{itemize}
    
\item {Actions:} at each timestep within an episode, Bob’s agent selects one of two possible actions. The first action corresponds to choosing a test angle $\theta_2$, which is applied to one of the remaining quantum state copies, followed by measurement. This interaction produces a measurement outcome that is returned to the agent as part of the observation. Each such action consumes one quantum state copy and incurs a fixed penalty. The second action corresponds to termination, in which the agent decides to stop further interaction and output a final estimate of the hidden classical bit. Upon termination, the episode ends and a terminal reward of $+1$ or $-1$ is assigned depending on whether the estimate is correct, with accumulated penalties subtracted from the total episode reward, as illustrated in equation \ref{return}.

\item {Observations:} at each timestep, following the selection of a test-angle action, the environment returns an observation vector to the agent. The observation captures both the agent’s interaction choice and the resulting measurement statistics, and is defined as
\begin{equation}\label{observation}
\mathbf{o}_t = \big[ \sin(\phi_2^{(t)}), \; \cos(\phi_2^{(t)}), \; P_t(1), \; t/N \big],
\end{equation}
where $\phi_2^{(t)}$ denotes the test angle selected by the agent at timestep $t$, $P_t(1)$ represents the estimated probability of obtaining measurement outcome $1$ based on the chosen $\phi_2^{(t)}$ , and $t/N$ provides a normalized measure of episode progress, with $N$ denoting the total number of available quantum state copies. The angular input $\phi_2^{(t)}$, is represented using its sine and cosine components rather than the raw angle value. Empirically, providing the raw test angle $\phi_2^{(t)}$ directly to the agent resulted in slower convergence and reduced evaluation accuracy, whereas the sine-cosine representation yielded more stable training and improved performance.
\end{itemize}

\section{Agent Architectures}\label{sec:agents}
The agent is trained to interact with the environment in order to infer the hidden classical bit based solely on the measurement outcomes it observes. While the measurement probabilities are governed by the mathematical expressions in Eq. \ref{probabilities}, these relationships are not provided explicitly to the agent. Instead, the agent learns an effective probing and stopping strategy through RL by observing the outcomes of its interactions. The agent learns to select test angles that yield informative measurement statistics and to decide when sufficient evidence has been collected to make a final guess. This requires implicitly capturing the relationship between probe angles and measurement outcomes for the two possible hidden bits, rather than explicitly minimizing the difference between $\phi_1$ and $\phi_2$.

To demonstrate the solvability of the proposed environment and to evaluate the behavior of different policy representations, we consider three classes of agents: a purely classical agent (C-agent), a deeper hybrid quantum–classical agent with a more expressive policy architecture (D-agent), and a simple hybrid quantum–classical agent with a lightweight policy representation (S-agent).  All agents are trained and evaluated in the same environment under identical conditions, including the use of the same random seeds.

\subsection{C-agent}
The classical agent employs a standard policy-gradient REINFORCE algorithm with a purely classical neural network policy that maps the observation vector to an action distribution over the available actions. The network takes as input the observation defined in Section~\ref{sec:environment} and outputs either a test-angle action or a termination action. This agent serves as a baseline for assessing the benefit of incorporating quantum components into the policy. 

The agent is trained by minimizing the REINFORCE loss function
\begin{equation}\label{classicalLoss}
L(\theta)
=
-\mathbb{E}\!\left[
\sum_{t=0}^{T-1}
\log \pi_\theta(a_t \mid o_t)\, \hat{G}_t 
\right],
\end{equation}

Here, $\hat{G}_t $ represents the variance-normalized return. In the proposed environment, the return at timestep $t$ corresponds to the remaining episodic return, with the initial return $G_0$ being equivalent to the episodic reward $R_{\mathrm{episode}}$.

\subsection{D-agent}
The first hybrid agent considered in this work follows a hybrid quantum-classical policy gradient approach with a baseline term subtracted from the return. The baseline is represented by a learned value function and is introduced to reduce variance and improve training stability. This formulation corresponds to the advantage actor-critic paradigm, where the policy network acts as the actor and the value function serves as the critic.

\begin{figure}
    \centering
    \includegraphics[width=1\linewidth]{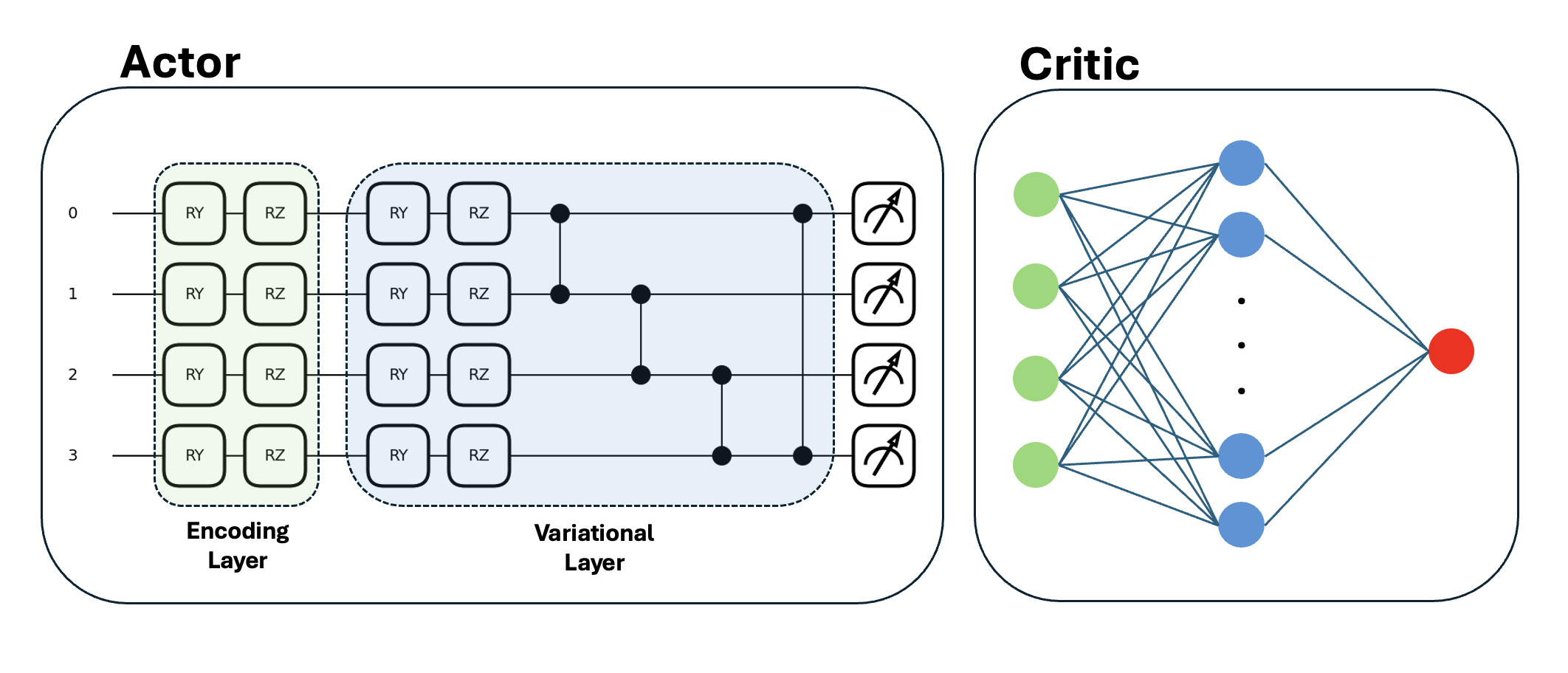}
    \caption{D-agent agent architecture}
    \label{fig:Hybrid1}
\end{figure}

In this agent, the actor is implemented as a parameterized quantum circuit, while the critic is realized as a classical neural network, as illustrated in Fig. \ref{fig:Hybrid1}. The critic estimates the state value and is used solely as a baseline during training, whereas action selection is entirely governed by the quantum policy.
The quantum actor adopts a hardware-efficient ansatz, which is known to provide strong expressivity with shallow circuit depth \cite{skolik2022quantum, jerbi2021parametrized}. The circuit is composed of alternating encoding and variational layers. The encoding layer embeds the observation into the quantum state using single-qubit rotation gates $R_y$ and $R_z$. The variational layer consists of trainable single-qubit rotations, again using $R_y$ and $R_z$. followed by entangling controlled-Z (CZ) gates. These alternating layers enable the circuit to learn expressive policies while remaining compatible with near-term quantum hardware constraints.

It is worth mentioning that the encoding layers are themselves parameterized and trainable, and are applied in alternation with variational layers, enabling a data re-uploading strategy that enhances the expressivity of the quantum policy.

The loss function is defined as follows: 

\begin{align}
y_t &= r_t + V_\psi(o_{t+1}), \\
A_t &= y_t - V_\psi(o_t), \\
\mathcal{L}_{\mathrm{actor}}(\theta) &=
-\sum_{t=0}^{T-1}\log\pi_\theta(a_t\mid o_t)\,A_t, \\
\mathcal{L}_{\mathrm{critic}}(\psi) &=
\frac{1}{2}\sum_{t=0}^{T-1}\big(y_t - V_\psi(o_t)\big)^2, \\
\mathcal{L}(\theta,\psi) &= \mathcal{L}_{\mathrm{actor}}(\theta) + \mathcal{L}_{\mathrm{critic}}(\psi).
\end{align}

Here, $V_\psi(o_t)$ denotes the critic’s estimate of the state-value.

\subsection{S-agent}
\begin{figure}
    \centering
    \includegraphics[width=1\linewidth]{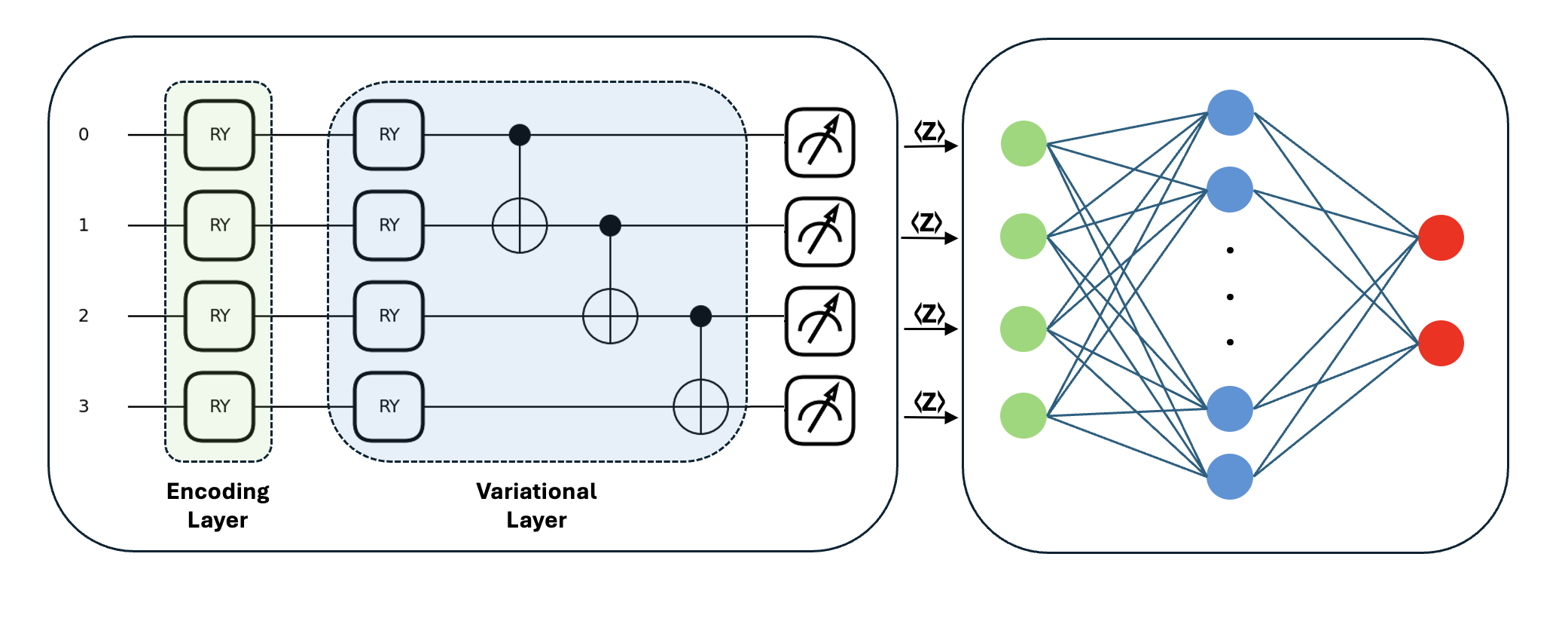}
    \caption{S-agent agent architecture}
    \label{fig:Hybrid2}
\end{figure}

The second hybrid agent, referred to as S-agent, adopts a simple, lightweight hybrid quantum–classical policy architecture, as shown in Fig. \ref{fig:Hybrid2}.  In contrast to D-agent, the quantum component is implemented as a shallow parameterized quantum circuit composed of a single encoding layer followed by a variational layer, both constructed using $R_y$ rotation gates. Unlike D-agent, the encoding layer is not repeated, and no data re-uploading strategy is employed, resulting in a significantly simpler and more resource-efficient quantum circuit.

Following the quantum circuit execution, expectation values of the Pauli-Z observables are measured. These expectation values are then passed to a classical neural network, which maps the extracted quantum features to an action distribution and produces the final policy output. 
In this architecture, both components are trainable: the parameters of the quantum circuit are optimized jointly with the weights of the classical neural network. 

The loss function used for training S-agent follows the same REINFORCE formulation as the C-agent (Eq.~\ref{classicalLoss}). The difference lies solely in the policy representation, where the observation is first processed by a quantum circuit before being passed to the classical network.

\section{Experimental Setup}\label{sec:experimentalsetup}
This section details the hyperparameters used for each of the three agents (C-agent, D-agent and S-agent). We also describe the two environment scenarios considered in our experiments, which impose different levels of constraint on the agents’ use of quantum state copies.
\subsection{Agents}
All agents are trained and evaluated within the same environment configuration to ensure a fair comparison. The hyperparameters used for training the different agents are summarized in Table~\ref{tab:hyperparams}. For D-agent, different learning rates are used for the encoding, variational, and output parameters of the quantum actor, while the critic is trained with a single learning rate.

\begin{table*}[t]
\centering
\caption{Experimental hyperparameters for all evaluated agents}
\label{tab:hyperparams}
\renewcommand{\arraystretch}{1.2}
\setlength{\tabcolsep}{5pt}
\begin{tabular}{l c c c c c c c c}
\hline
\textbf{Agent} 
& \textbf{Quantum Depth} 
& \textbf{Classical Depth} 
& \textbf{LR (Quantum Actor)} 
& \textbf{LR (Classical Policy)} 
& \textbf{LR (Critic)} 
& \textbf{Training Batch} 
& \textbf{Eval. Episodes}
& $\boldsymbol{\gamma}$ \\
\hline

C-agent 
& -- 
& 1 (128 neurons)
& -- 
& 0.01
& -- 
& 30 
& 1000
& 1 \\

D-agent 
& 4 
& 1 (64 neurons) 
& $\{0.01,\;0.01,\;0.005\}$ 
& -- 
& 0.001 
& 30 
& 1000
& 1 \\

S-agent 
& 2 
& 1 (128 neurons)
& -- 
& 0.01
& -- 
& 30 
& 1000
& 1 \\

\hline
\end{tabular}
\end{table*}

\subsection{Environment}
We consider two environment scenarios that differ in the interaction penalty associated with consuming quantum copies. In both scenarios, the agent is given access to a maximum of $N = 10$ quantum copies per hidden bit-angle pair, and each interaction corresponds to the consumption of one copy. An episode terminates when the agent outputs a prediction or when the available copies are exhausted. The scenarios differ in the amount of penalty incurred while fixing all other environment parameters. 
\begin{itemize}
\item High-penalty scenario: in this scenario, each interaction incurs a penalty of $X = 0.5$, enforcing strict constraints on quantum resource usage. 
\item Low-penalty scenario: in this scenario, the interaction penalty is reduced to $X = 0.05$, allowing greater exploratory behavior while still discouraging excessive interaction.
\end{itemize}

For both settings, the number of measurement shots is fixed at $\text{shots} = 4$ per interaction. This choice reflects a trade-off between resource efficiency and accuracy, providing sufficient reliability while limiting quantum measurement overhead. The learning behavior of all agents is evaluated under these two penalty regimes, with performance metrics reported in the following section.

\section{Evaluation Metrics}\label{sec:evaluation}
The main objective of the proposed environment is for an agent to correctly infer and extract the hidden information embedded in the quantum circuit parameters within a limited number of interactions. Accordingly, the performance of each agent is evaluated based on its ability to successfully solve the challenge-response task. 
Accuracy is defined as the fraction of episodes in which the agent correctly identifies the hidden bit upon termination. This metric directly reflects the agent’s ability to infer the encoded information under the interaction constraints imposed by the environment.  Specifically, accuracy is computed as:
\begin{equation}
\mathrm{Accuracy}
=
\frac{1}{N_{\mathrm{ep}}}
\sum_{i=1}^{N_{\mathrm{ep}}}
\mathbb{I}\!\left( \hat{b}_i = b_i \right),
\end{equation}
where $b_i$ denotes the true hidden bit in episode $i$, $\hat{b}_i$ is the agent’s predicted bit at episode, $\mathbb{I}(\cdot)$ is the indicator function, and $N_{\mathrm{ep}}$ is the number of evaluation episodes ($N_{\mathrm{ep}} = 1000$ in our experiments).

In addition to accuracy, we report the following metrics:
\begin{itemize}
    \item The average number of interactions (or consumption of quantum copies) per episode.
    \item The confusion matrix for each agent provides a detailed breakdown of correct and incorrect predictions across the two possible hidden bit values.
\end{itemize}

\section{Results}\label{sec:results}

Our goal is to evaluate the solvability of the proposed quantum RL environment and to compare the learning behavior of different agent realizations under identical conditions.
We analyze agent training/evaluation performance under the two environment scenarios introduced in Section~\ref{sec:experimentalsetup}, corresponding to low and high interaction penalties. For each scenario, we report the learning behavior during training as well as the final evaluation performance. This includes accuracy, average number of interactions per episode, and confusion matrix analysis. To assess robustness and resource efficiency, the trained agents are evaluated in a constrained evaluation setting in which only two quantum state copies are available per episode. Evaluation is performed over three independent random seeds, and accuracy results are reported as the mean and standard deviation across evaluation runs.

\subsection{High-Penalty Environment}
\begin{figure}
    \centering
    \subfigure[C-agent]{\includegraphics[width=0.49\textwidth]{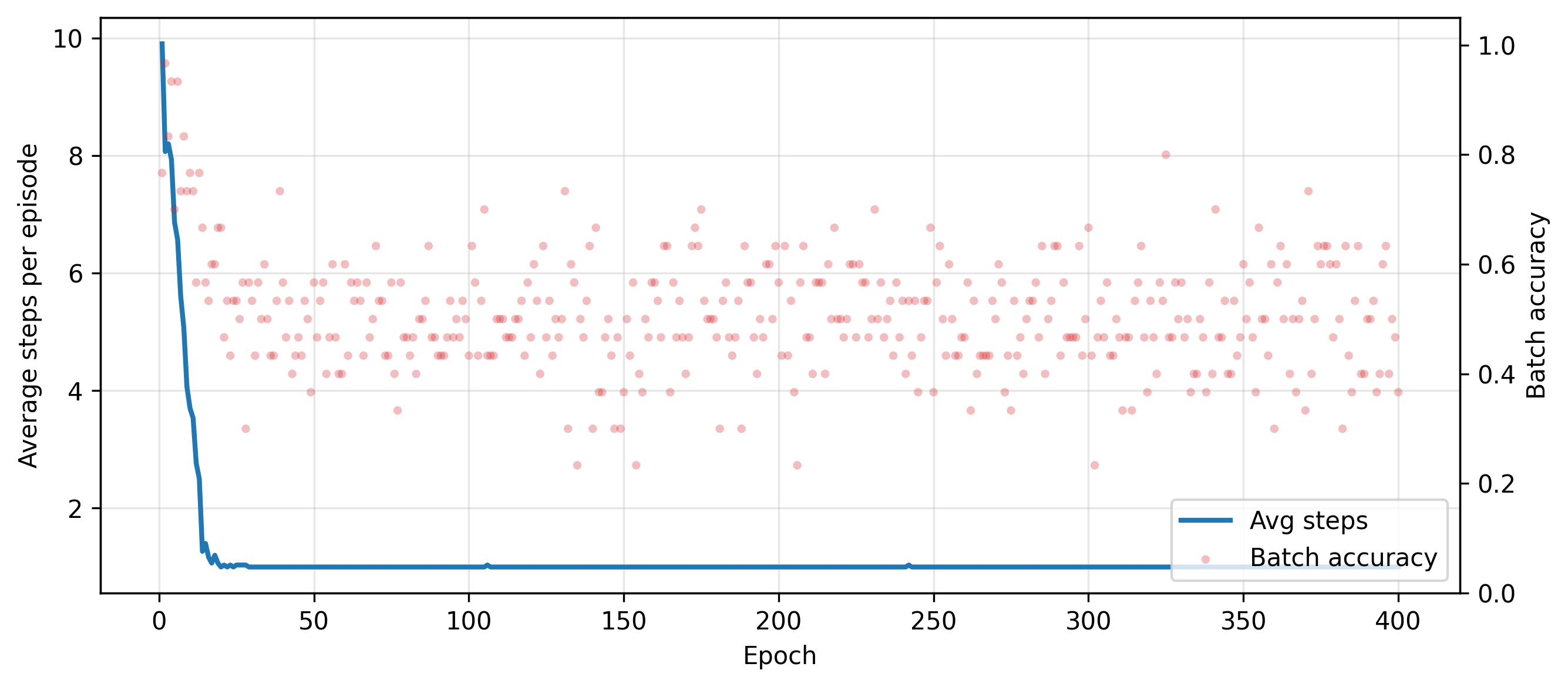}}
    \subfigure[D-agent]{\includegraphics[width=0.49\textwidth]{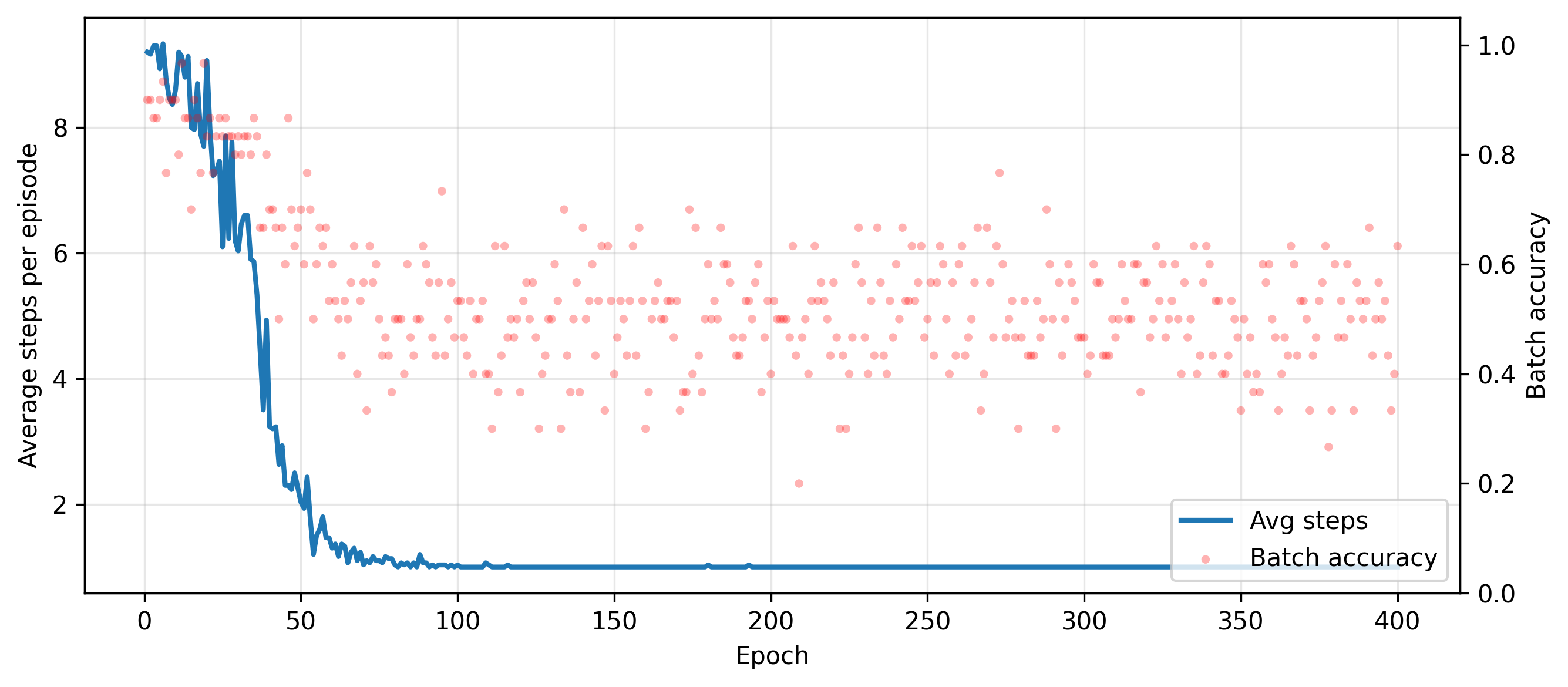}} 
    \subfigure[S-agent]{\includegraphics[width=0.49\textwidth]{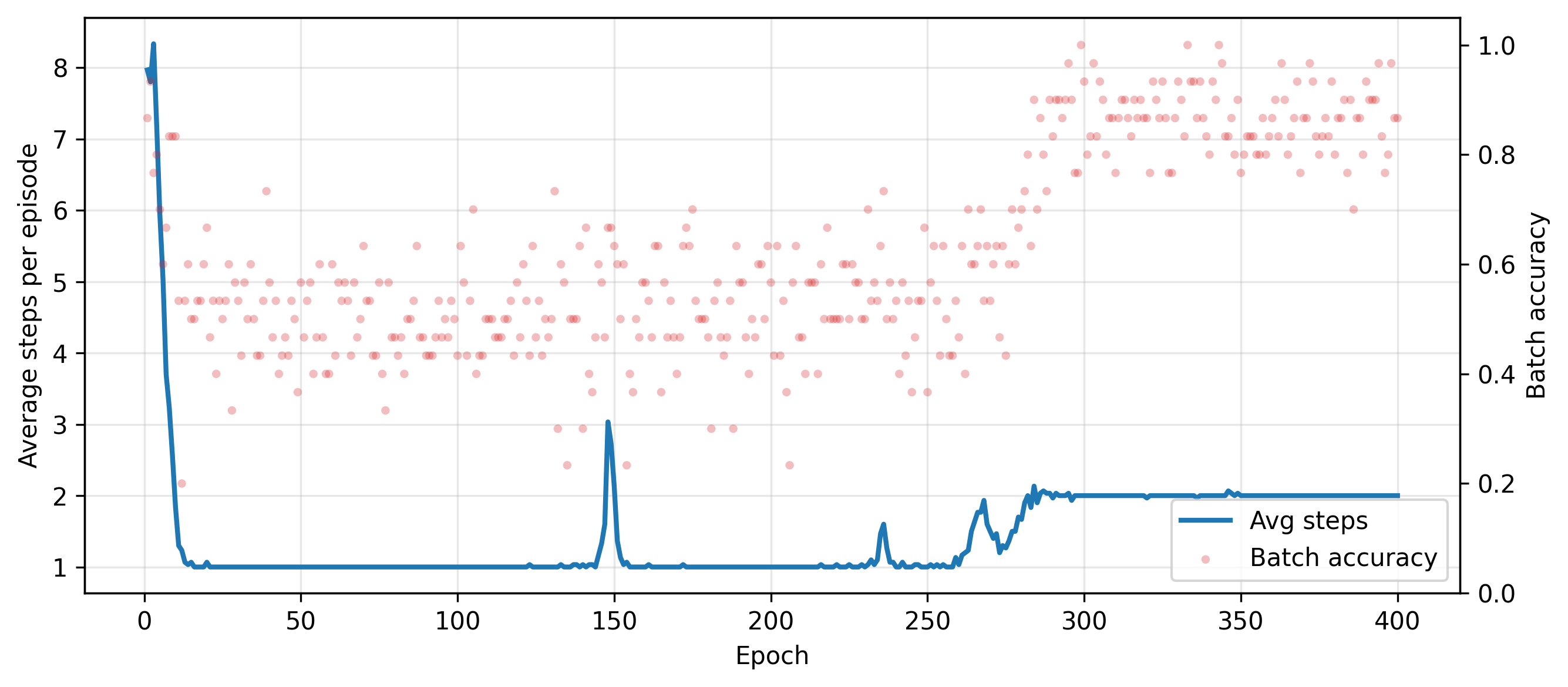}} 
    \vspace{-10pt}
    \caption{Training under high-penalty environment ($X=0.5$).
The solid curve shows the average number of interactions per episode, while the scatter points indicate batch accuracy.}
    \label{fig:training_0.5}
\end{figure}

We first consider a high-penalty environment, where each interaction incurs a larger cost ($X=0.5$), strongly discouraging excessive probing (i.e., consumption of quantum copies). Fig. \ref{fig:training_0.5} illustrates the evolution of the average number of interactions per episode and the corresponding batch accuracy during training. Table \ref{tab:accuracy_0.5} shows the accuracy across all three agents. 

As shown in Fig.~\ref{fig:training_0.5}(a), although the C-agent initially has access to up to ten quantum copies per episode $N=10$, it rapidly adapts to the high step penalty by aggressively reducing its interactions. Within the first few epochs, the average number of steps per episode drops to nearly one and remains at this level throughout training. This behavior corresponds to testing a single angle on a single quantum copy. While this strategy minimizes the incurred cost, it provides insufficient information to distinguish between the probability distributions associated with the two hidden bits. As a result, the batch accuracy remains low and fluctuates around $0.5$, indicating near-random inference under strict probing constraints.

D-agent, shown in Fig.~\ref{fig:training_0.5}(b), exhibits behavior similar to the C-agent under high-penalty, adapting to the step cost by progressively reducing the number of interactions per episode. Unlike the abrupt collapse observed for the classical baseline, the reduction in steps occurs more gradually, indicating a smoother trade-off between exploration and cost minimization. However, this gradual decrease in interactions is accompanied by a corresponding gradual decline in accuracy. As training progresses and the agent converges toward single-angle  probing, the accuracy remains comparable to that of the C-agent, suggesting that the inclusion of a deeper quantum actor alone does not improve inference performance in this highly restrictive setting.

S-agent, shown in Fig.~\ref{fig:training_0.5}(c), demonstrates the most effective trade-off under the high-penalty regime. During the early stages of training, the agent initially operates with very low interaction counts, close to single-angle probing, before adjusting its strategy. As training progresses, the agent converges to a stable policy with approximately $T=2$ interactions per episode while maintaining consistently high accuracy. Unlike the C-agent and D-agent, S-agent does not remain trapped in single-angle probing, indicating that the lightweight quantum circuit enables more informative measurements from a small number of interactions. This behavior suggests that a compact quantum policy can extract sufficient information for reliable inference even under severe resource constraints.

\begin{figure*}
    \centering
    \subfigure[C-agent]{\includegraphics[width=0.29\textwidth]{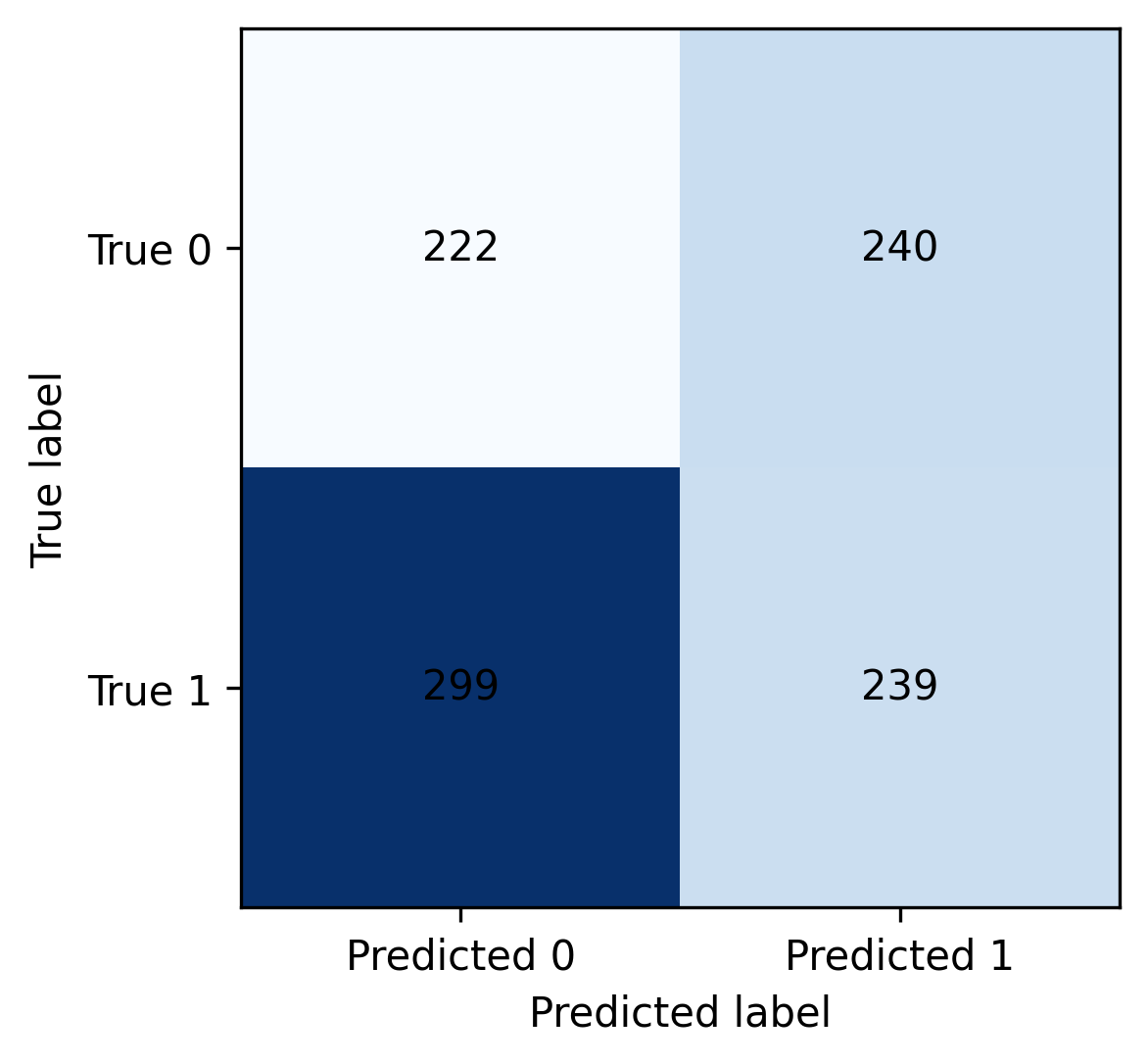}} 
    \subfigure[D-agent]{\includegraphics[width=0.29\textwidth]{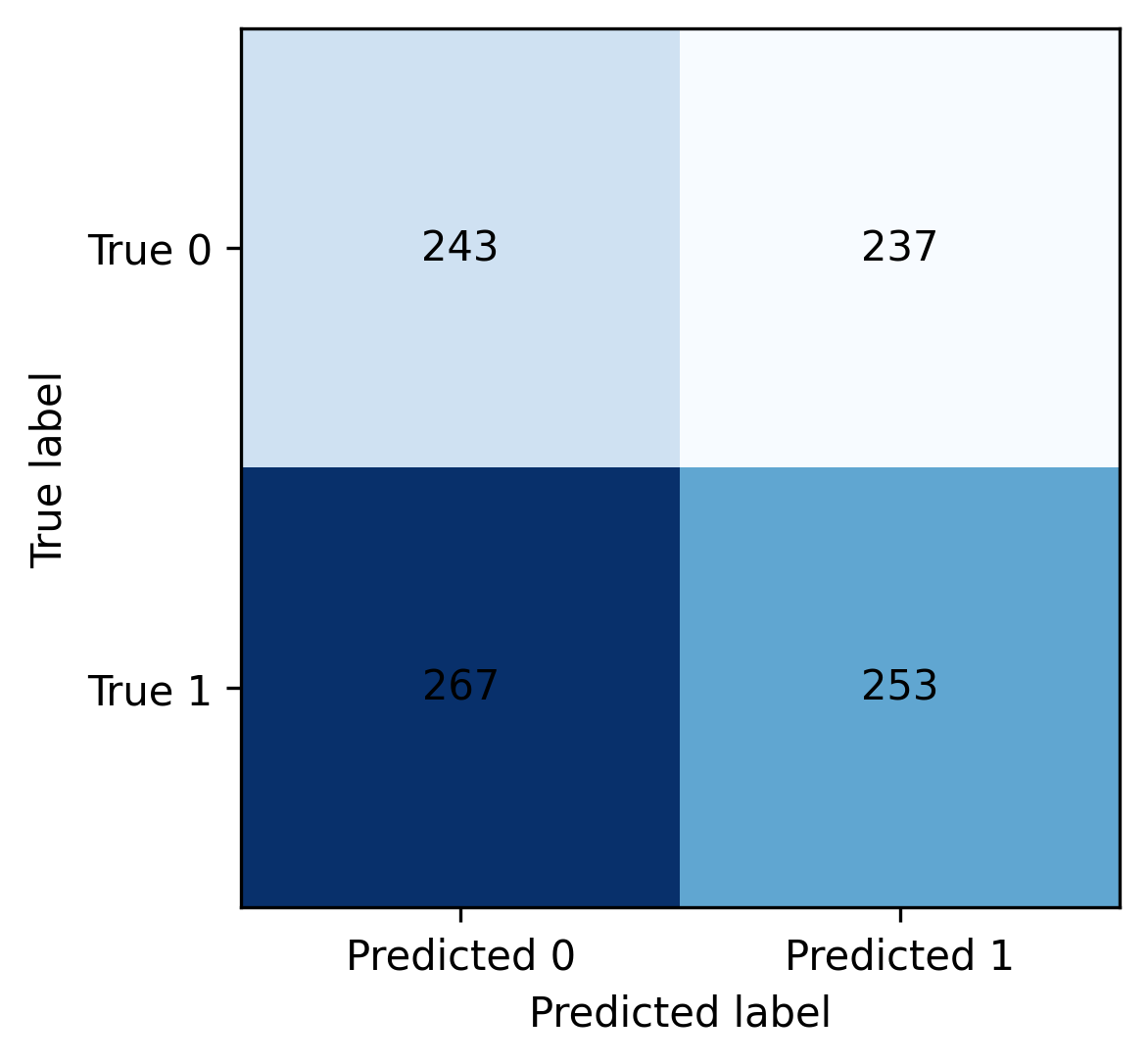}} 
    \subfigure[S-agent]{\includegraphics[width=0.29\textwidth]{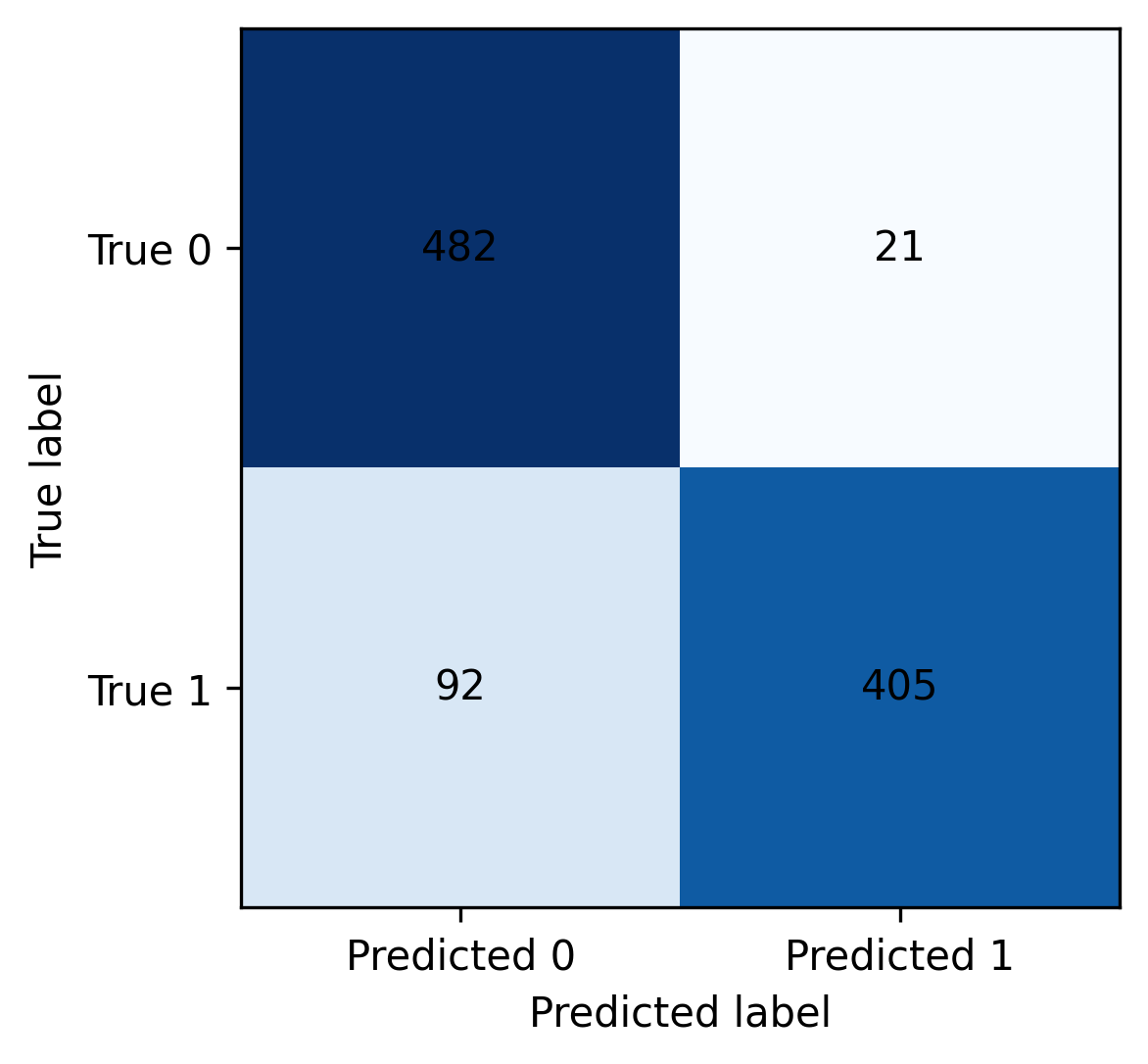}} 
    \vspace{-10pt}
    \caption{Confusion matrices under the high-penalty environment ($X=0.5$).}
    \label{fig:confusion_0.5}
\end{figure*}

Fig.~\ref{fig:confusion_0.5} reports the confusion matrices for the three agents, showing that the C-agent and D-agent exhibit similar classification behavior, while S-agent achieves more accurate and balanced predictions.

\begin{table}[t]
\centering
\caption{Evaluation accuracy under high-penalty environment}
\label{tab:accuracy_0.5}
\begin{tabular}{lc}
\hline
\textbf{Agent} & \textbf{Accuracy (\%)} \\
\hline
C-agent  & $0.49 \pm 0.03$ \\
D-agent  & $0.49 \pm 0.03$
 \\
S-agent  & $0.880 \pm 0.010
$ \\
\hline
\end{tabular}
\end{table}

\subsection{Low-Penalty Environment}
Next, we consider a low-penalty environment, where each interaction incurs a smaller cost ($X=0.05$), allowing greater exploratory behavior during training. We follow a similar assessment as in the previous environment. Table \ref{tab:accuracy_0.05} shows the accuracy across all three agents. 

In this setting, all agents initially achieve high accuracy from the early stages of training. This behavior is expected, as the low step penalty permits extensive probing: agents begin by utilizing nearly the maximum number of available quantum copies, effectively testing many angles per episode. Under such conditions, inferring the hidden bit becomes relatively straightforward, as the agent can rely on majority voting across multiple measurements rather than learning an efficient probing strategy. The central learning challenge here is therefore not achieving high accuracy, but maintaining it while reducing the number of interactions.

\begin{figure}
    \centering
    \subfigure[C-agent]{\includegraphics[width=0.49\textwidth]{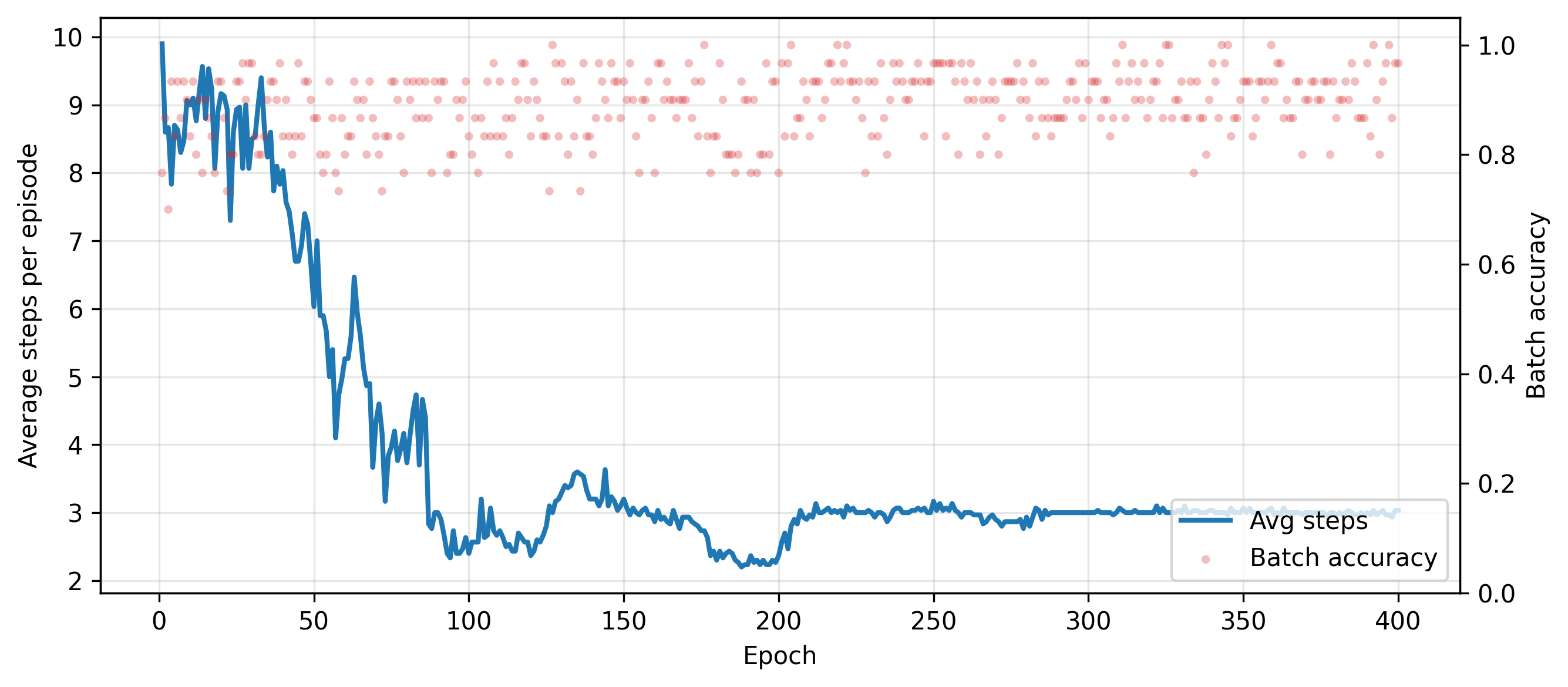}} 
    \subfigure[D-agent]{\includegraphics[width=0.49\textwidth]{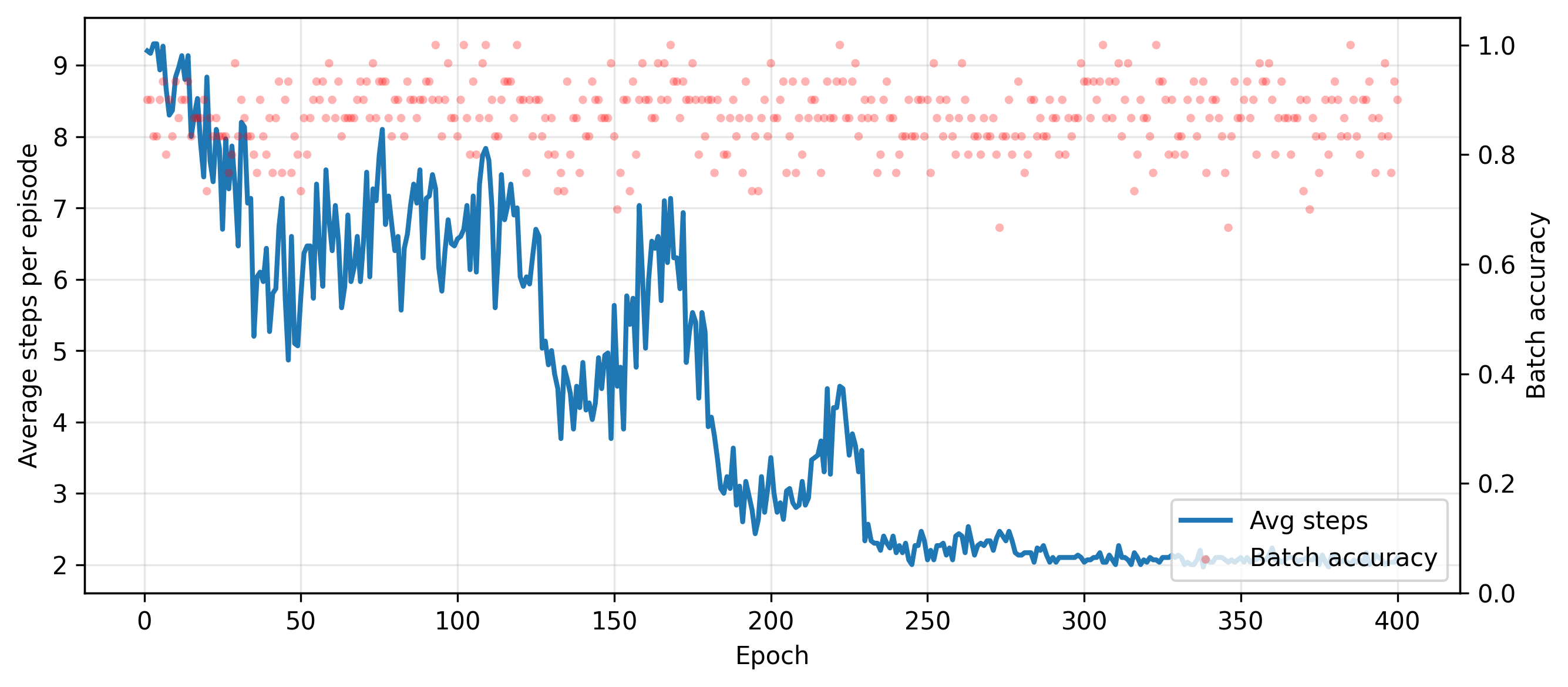}} 
    \subfigure[S-agent]{\includegraphics[width=0.49\textwidth]{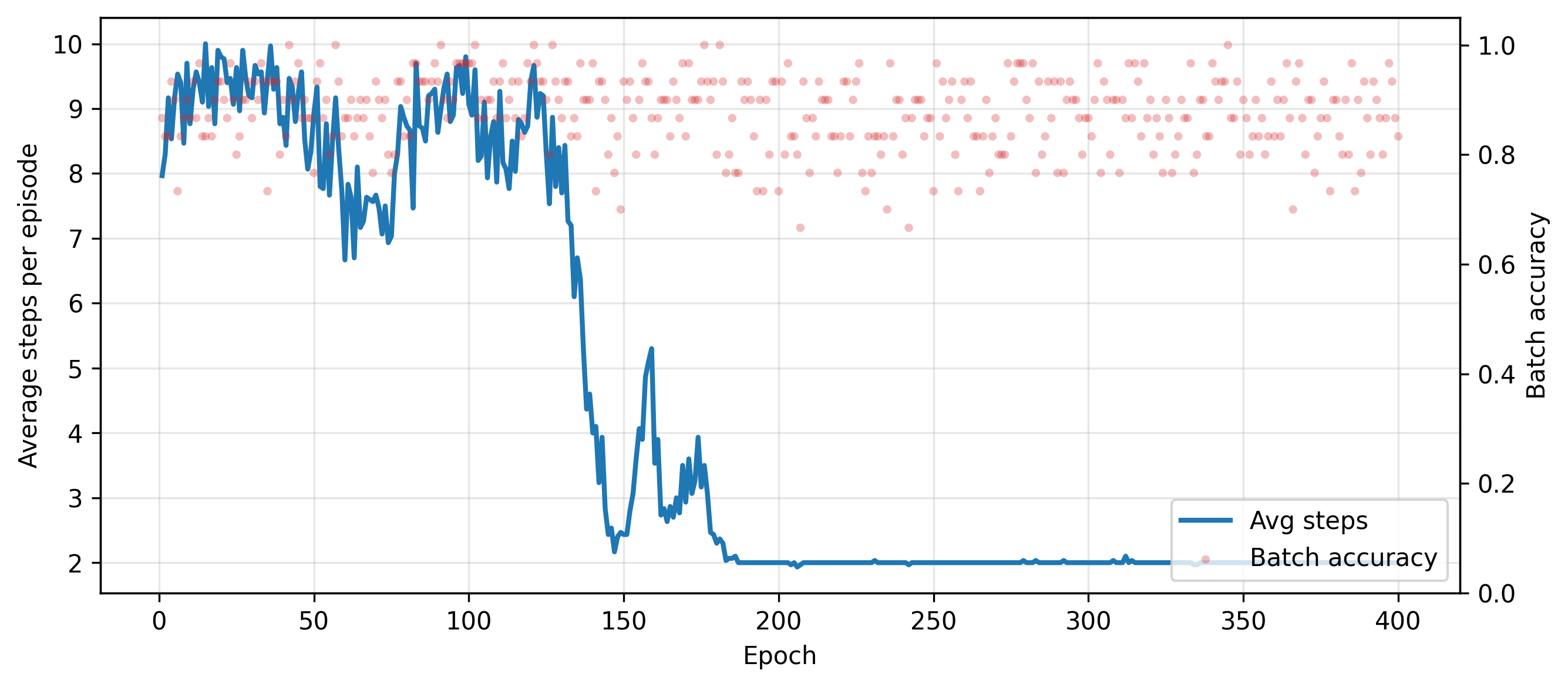}} 
    \vspace{-10pt}
    \caption{Training under low-penalty environment ($X=0.05$).
The solid curve shows the average number of interactions per episode, while the scatter points indicate batch accuracy.}
    \label{fig:training_0.05}
\end{figure}

Fig.~\ref{fig:training_0.05}(a) illustrates the training behavior of the C-agent. Unlike the high-penalty environment, the agent is not forced to immediately minimize its interactions. Instead, it gradually reduces the average number of steps per episode, eventually converging to approximately $T=3$ steps while preserving high accuracy throughout training. This gradual reduction reflects the agent’s ability to trade excess probing for more informed angle selection without sacrificing inference performance.

The training behavior of the D-agent agent is shown in Fig.~\ref{fig:training_0.05}(b). Similarly to the C-agent, D-agent begins with high probing and high accuracy, but converges more efficiently to a lower interaction regime. In particular, the agent stabilizes around $T=2$  steps per episode while consistently maintaining high accuracy. Compared to the classical baseline, D-agent achieves reliable inference using fewer quantum copies, indicating improved learning efficiency.

Fig.~\ref{fig:training_0.05}(c) Shows the training behavior of the S-agent agent. A similar trend to D-agent is observed; however, S-agent converges more rapidly, with stabilization occurring around epoch 190. As with D-agent, the agent ultimately converges to $T=2$ interactions per episode while maintaining high accuracy, demonstrating that effective inference can be achieved with minimal probing.

\begin{figure*}
    \centering
    \subfigure[C-agent]{\includegraphics[width=0.29\textwidth]{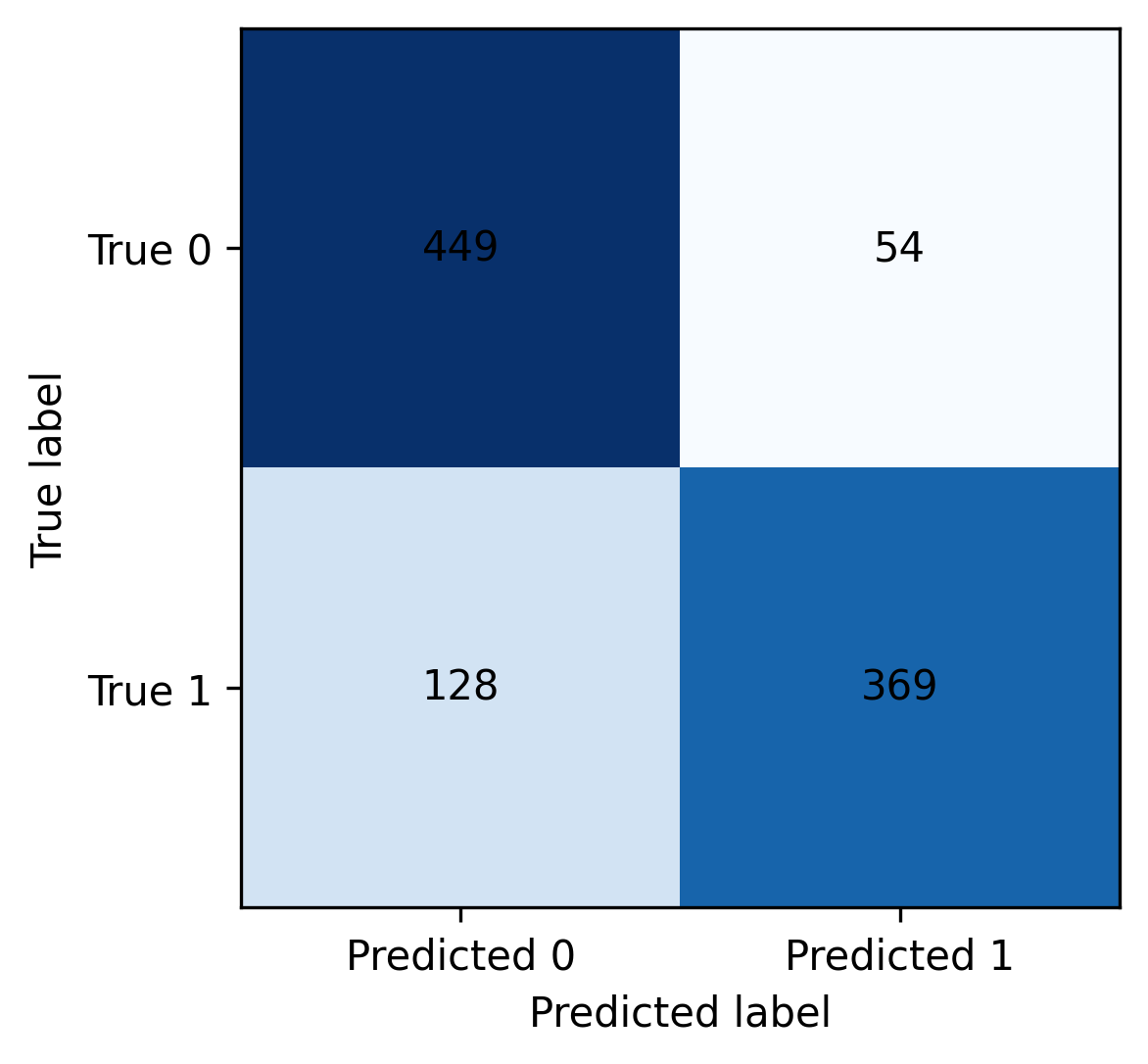}} 
    \subfigure[D-agent]{\includegraphics[width=0.29\textwidth]{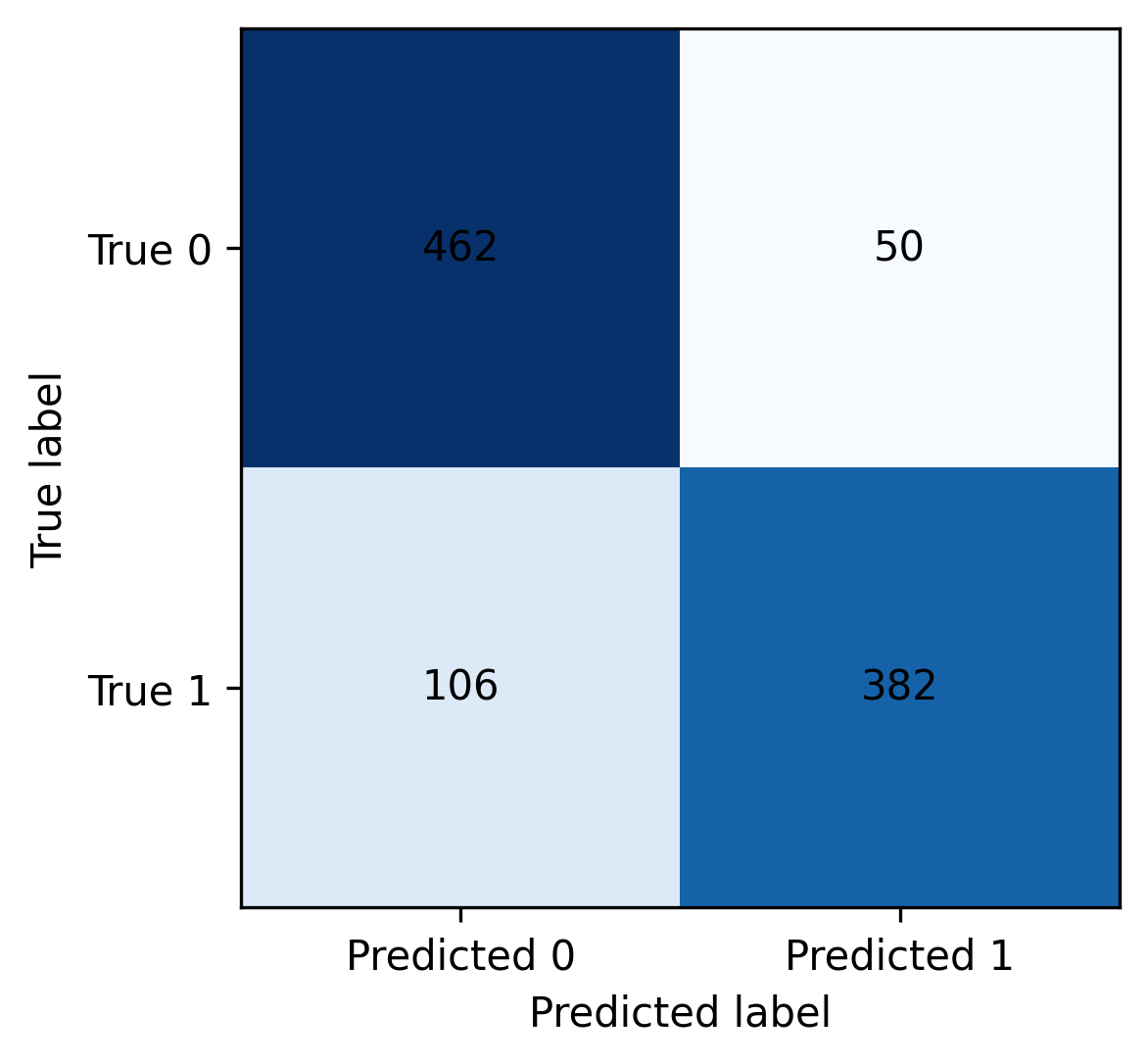}} 
    \subfigure[S-agent]{\includegraphics[width=0.29\textwidth]{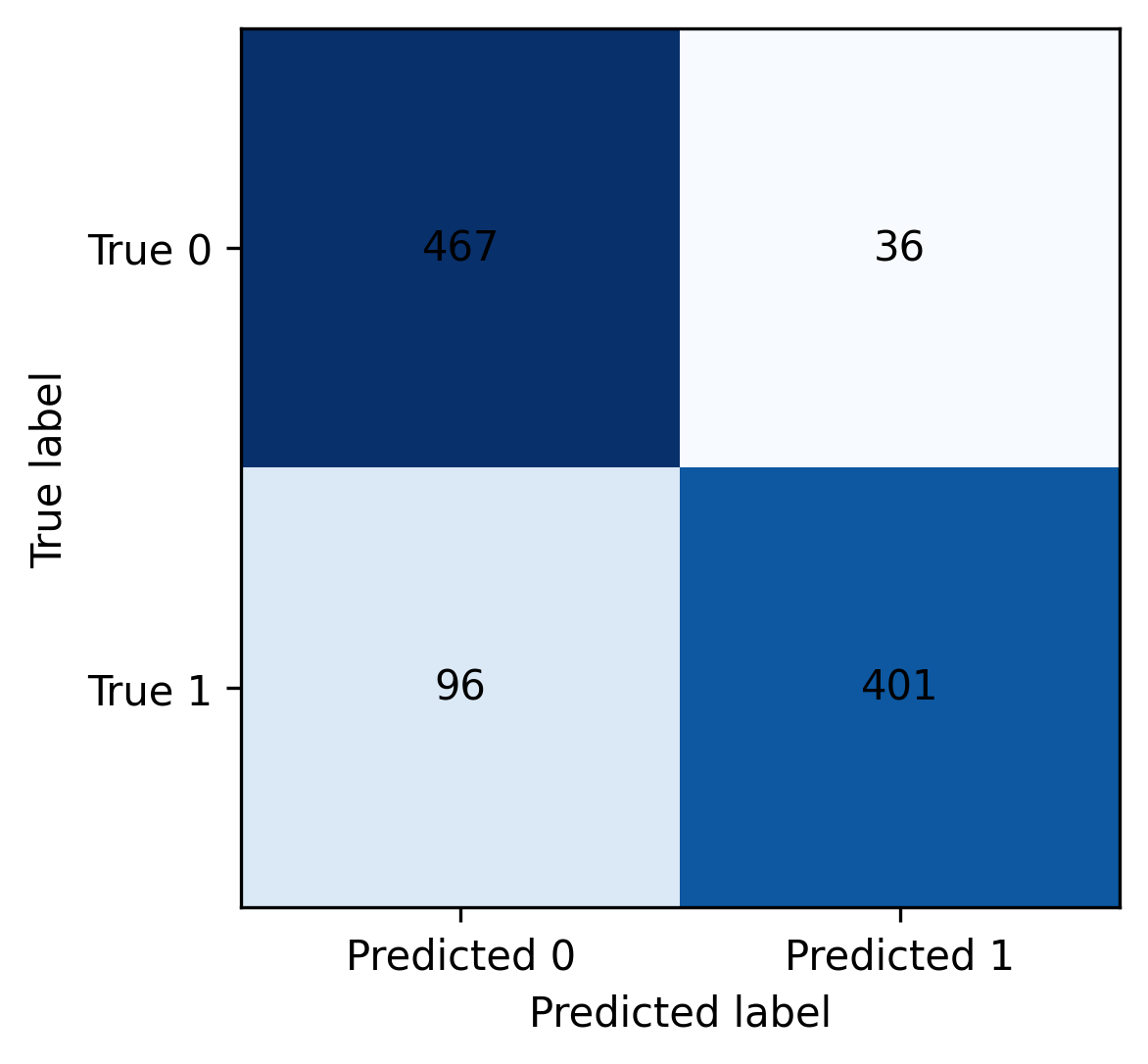}} 
    \vspace{-10pt}
    \caption{Confusion matrices under the low-penalty environment ($X=0.05$).}
    \label{fig:confusion_0.05}
\end{figure*}

From these results, we conclude that both D-agent and S-agent outperform the C-agent, as they require fewer quantum copies to make a correct inference while achieving higher accuracy, with S-agent achieving the highest overall accuracy.. This demonstrates that hybrid agents are more resource-efficient in the low-penalty setting. This trend is further confirmed by the confusion matrices in Fig.~\ref{fig:confusion_0.05}, which show that both D-agent and S-agent achieve more balanced and reliable predictions across the two hidden bit values compared to the C-agent.

We note that S-agent employs a simpler quantum circuit than D-agent. In scenarios where circuit complexity is a critical consideration, S-agent offers a better trade-off by achieving higher accuracy than D-agent while requiring fewer quantum states and reduced circuit complexity, and thus lower implementation cost. 
This consistency highlights S-agent as a robust and resource-efficient policy across varying interaction costs.

\begin{table}[t]
\centering
\caption{Evaluation accuracy under low-penalty environment}
\label{tab:accuracy_0.05}
\begin{tabular}{lc}
\hline
\textbf{Agent} & \textbf{Accuracy (\%)} \\
\hline
C-agent  & $82.0 \pm 1.0$ \\
D-agent  & $85.8 \pm 1.4$
 \\
S-agent  & $87.2 \pm 1.0$ \\
\hline
\end{tabular}
\end{table}

\subsection{Robustness to Noise}
Alice sends the prepared qubits to Bob, who attempts to extract the hidden bits through adaptive measurements. In the ideal case, the measurement statistics follow Eq.~\ref{probabilities}. However, in practical quantum communication scenarios, transmitted quantum states are affected by noise arising from imperfect quantum channels, which in turn alters the observed measurement probabilities. To evaluate the robustness of the proposed agents under such realistic conditions, we assess their performance in the presence of common quantum noise channels. In this work, agents are trained in a noiseless environment and evaluated under noisy conditions. This evaluation protocol tests the agents’ ability to generalize to unseen noise without explicitly adapting their policies during training, thereby assessing robustness. We consider three representative quantum noise models:
\begin{itemize}
    \item \text{Bit-flip noise}: in this model, a qubit is flipped by the Pauli-$X$ operator, $|0\rangle \leftrightarrow |1\rangle$, with probability $p$, and left unchanged with probability $1-p$.
    \item \text{Depolarizing noise}: in this model,  the qubit’s state is replaced by a maximally mixed state with probability $p$.
    \item \text{Amplitude damping noise}: in this mode, a qubit relaxes from the excited state $|1\rangle$ to the ground state $|0\rangle$ with probability $p$, while $|0\rangle$ remains unchanged.
\end{itemize}

\begin{figure*}
    \centering
    \subfigure[Bit-flip noise]{\includegraphics[width=0.29\textwidth]{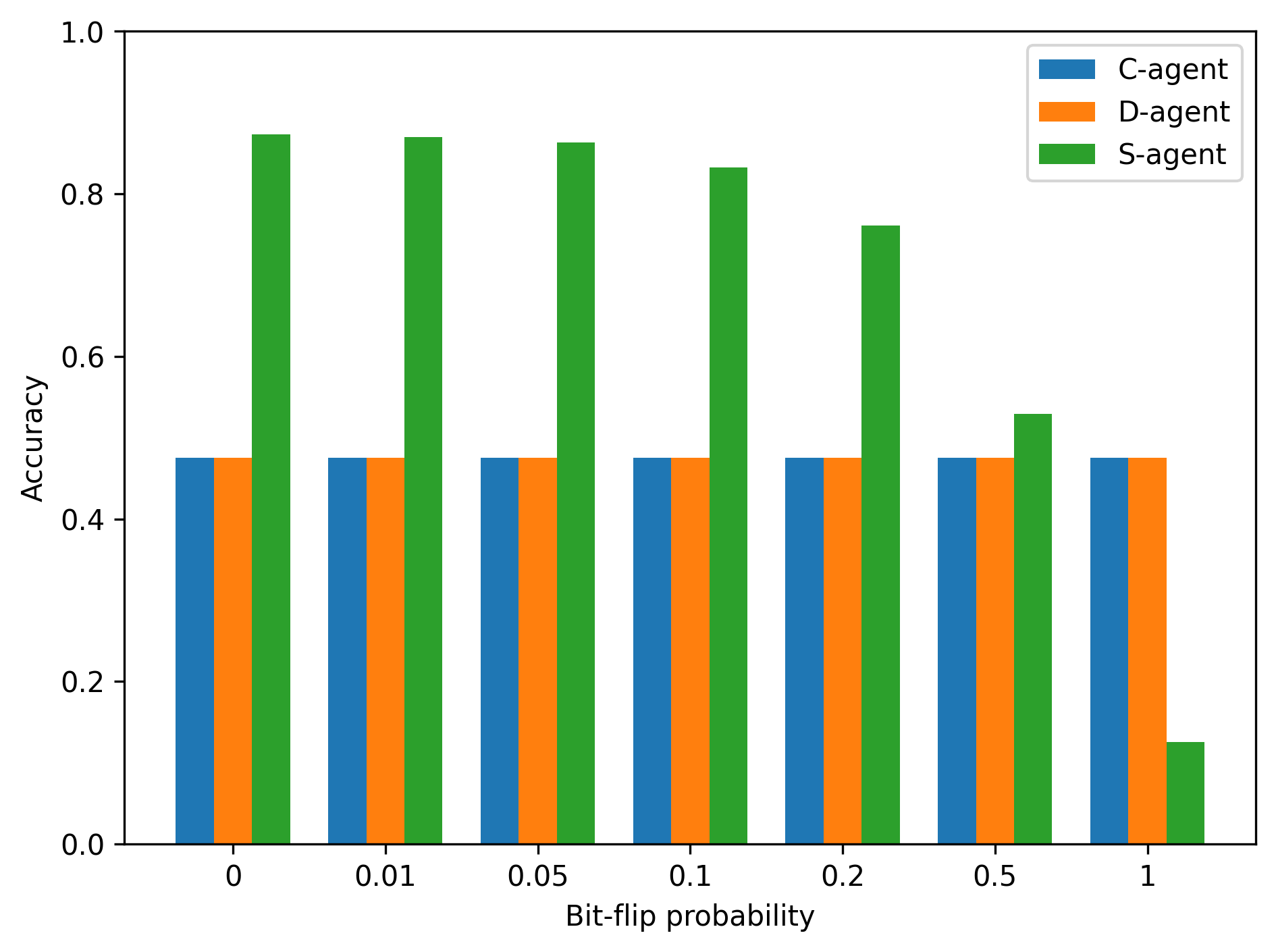}} 
    \subfigure[Depolarization noise]{\includegraphics[width=0.29\textwidth]{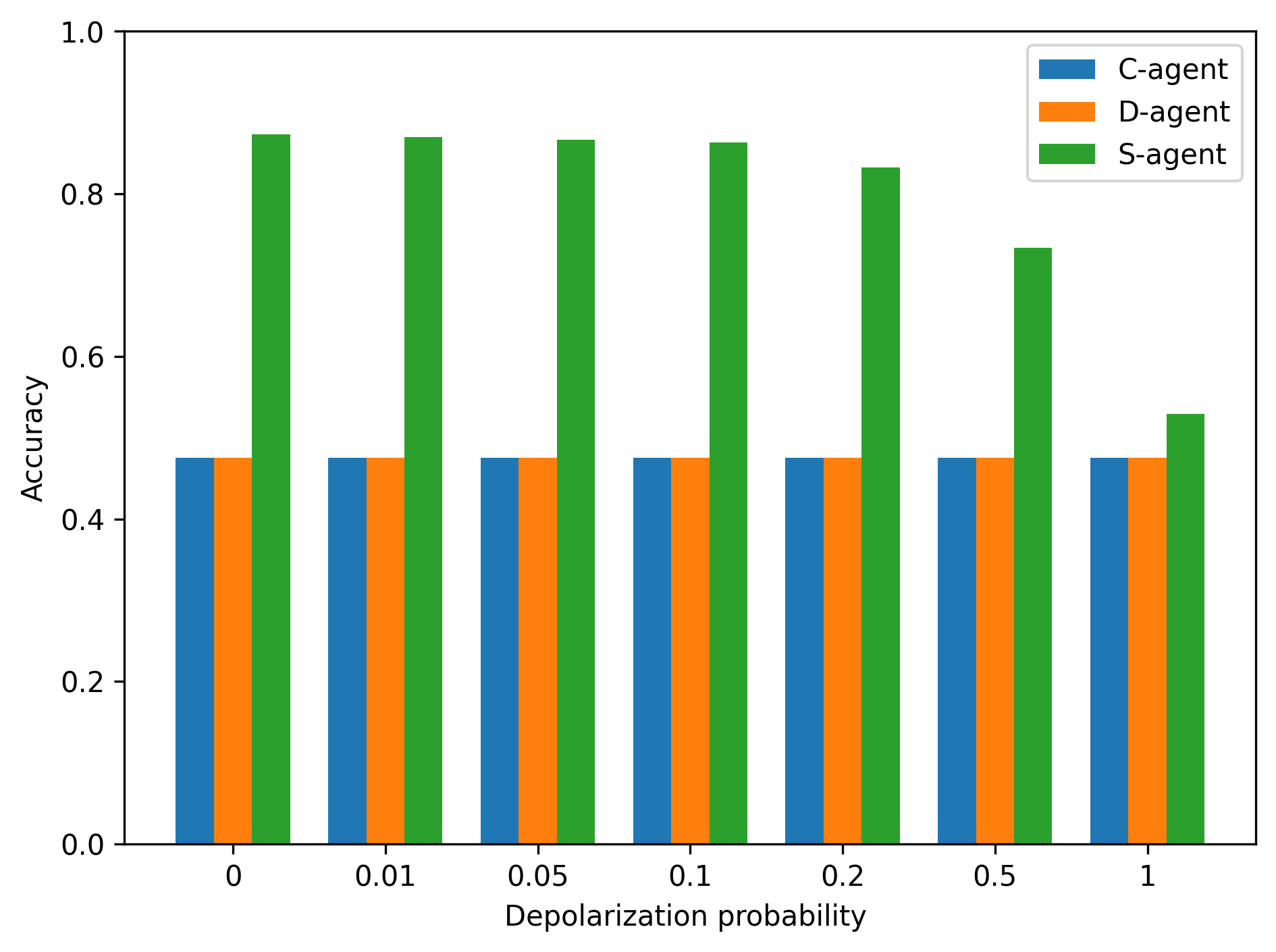}} 
    \subfigure[Amplitude damping noise]{\includegraphics[width=0.29\textwidth]{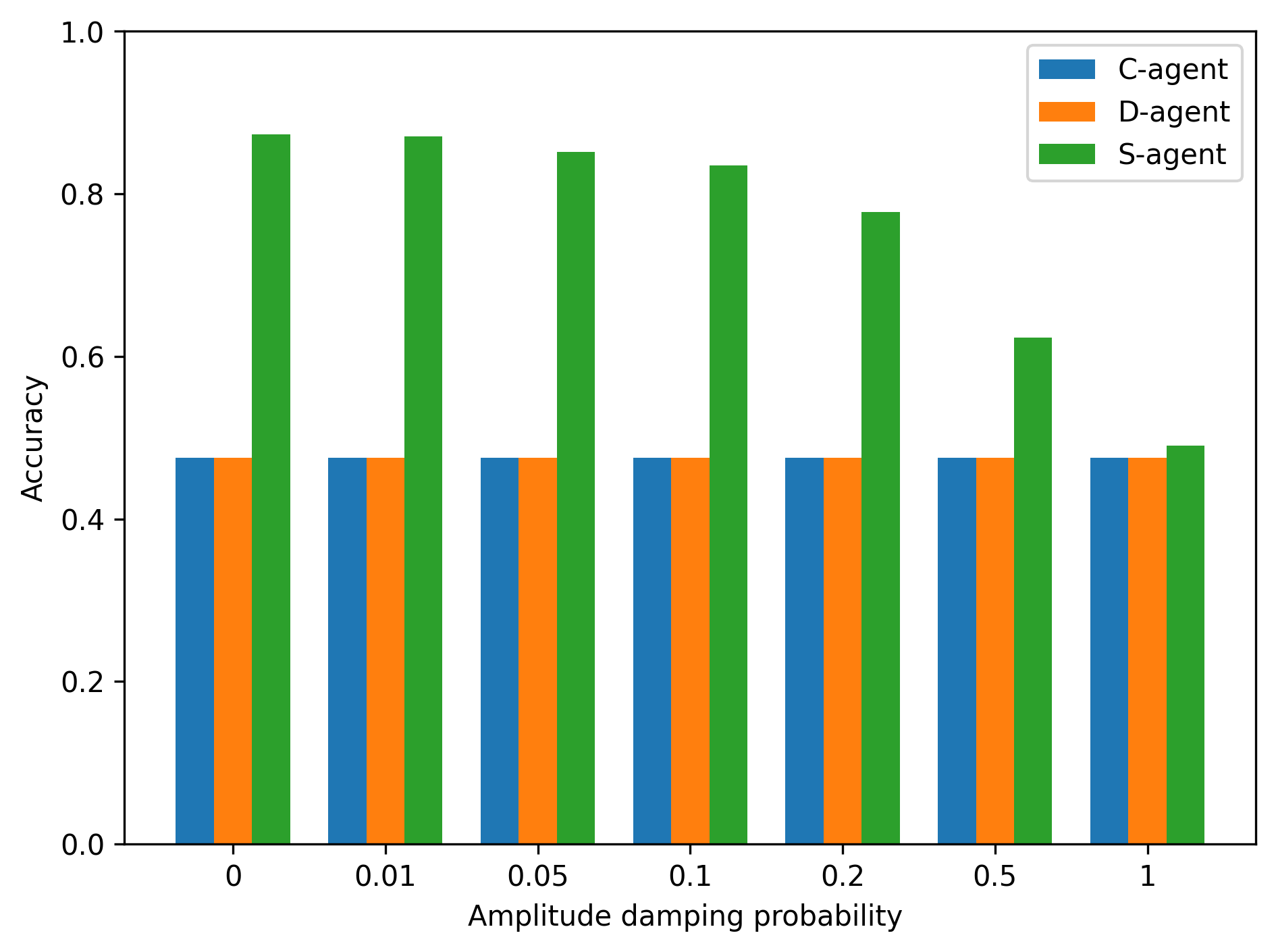}} 
    \vspace{-10pt}
    \caption{Accuracy across different noise models for agents trained on high penalty ($X=0.5$).
}
    \label{fig:noise_0.5}
\end{figure*}

Fig.~\ref{fig:noise_0.5} reports the classification accuracy of all agents under different quantum noise models when trained in the high-penalty environment ($X=0.5$).

Across all three noise models, bit-flip, depolarizing, and amplitude damping, the C-agent and D-agent agents exhibit random inference behavior, with accuracy remaining close to $0.5$ as the noise strength increases. This outcome is expected, as both agents already operate near randomly in the noiseless high-penalty setting due to their collapse toward minimal probing. Consequently, the introduction of channel noise does not change their performance.
In contrast, S-agent consistently demonstrates superior robustness across all noise models. Under low-to-moderate noise, S-agent maintains high accuracy relative to both the C-agent and D-agent, reflecting its ability to extract more informative measurements. Although all agents eventually converge toward $0.5$ accuracy at extreme noise strengths (e.g., probability of noise=1), S-agent degrades more gradually, preserving useful inference capability over a broader noise range. 

For the specific case of the bit-flip channel at probability of noise=1, the channel acts as a deterministic inversion. Every measured bit is flipped. Since S-agent learns a strong, structured policy in the noise-free case, this inversion causes it to consistently predict the opposite label. As a result, its accuracy approaches the complement of its noise-free performance ($1-0.873\approx0.127$) rather than converging to $0.5$. In contrast, the C-agent and D-agent do not learn a sharp policy and, therefore, degrade toward random guessing under extreme noise.

\begin{figure*}
    \centering
    \subfigure[Bit-flip noise]{\includegraphics[width=0.29\textwidth]{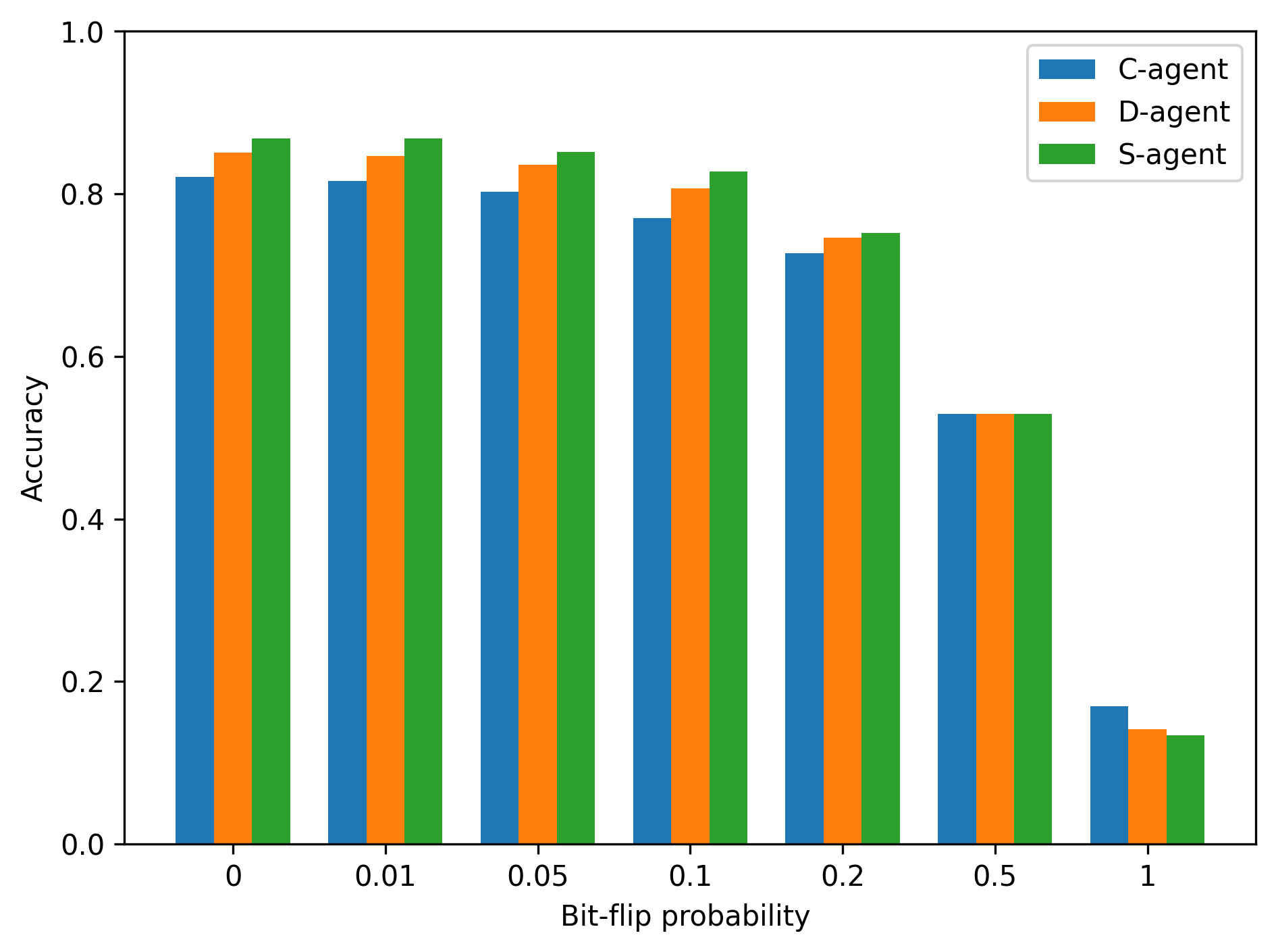}} 
    \subfigure[Depolarization noise]{\includegraphics[width=0.29\textwidth]{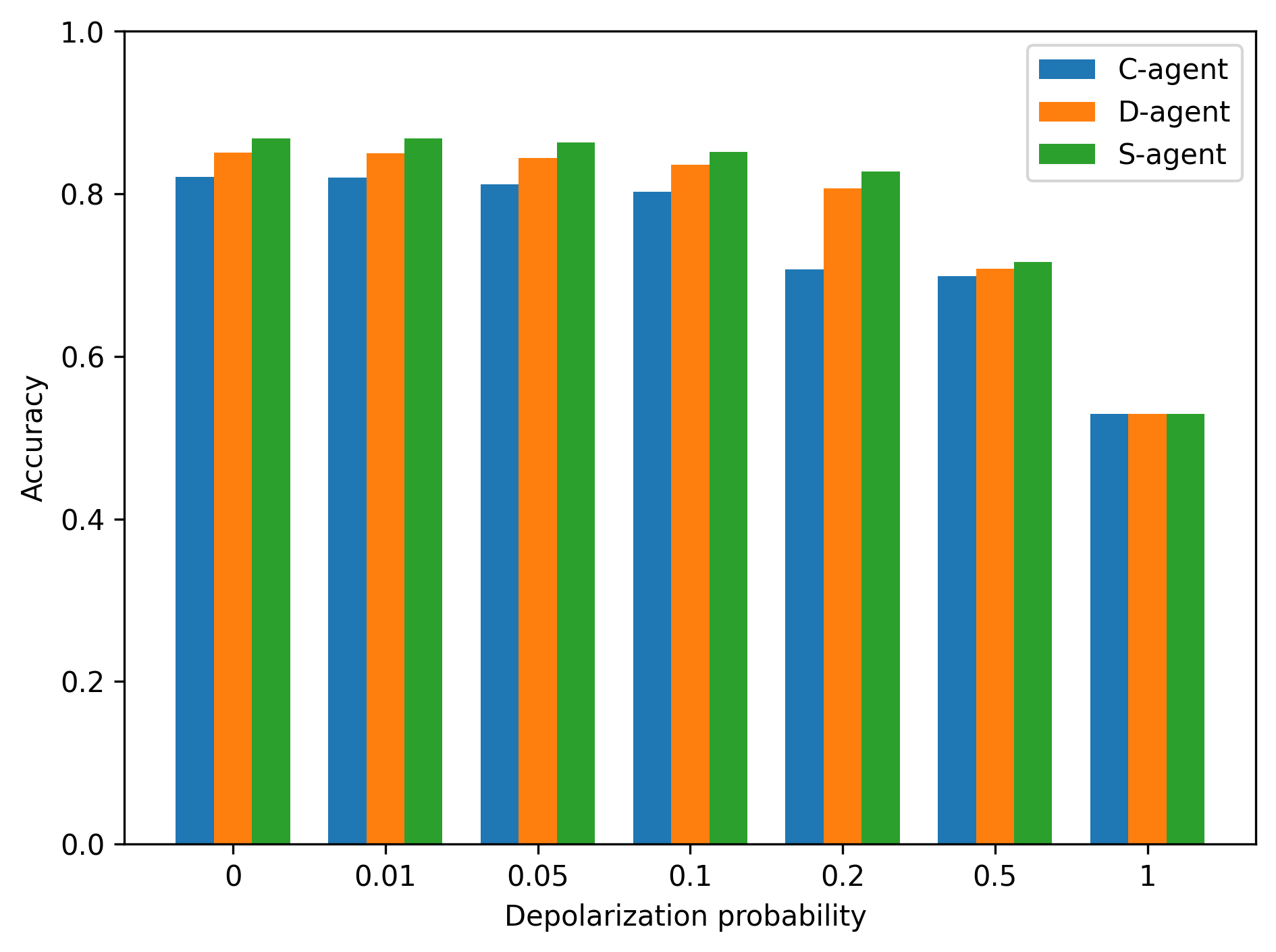}} 
    \subfigure[Amplitude damping noise]{\includegraphics[width=0.29\textwidth]{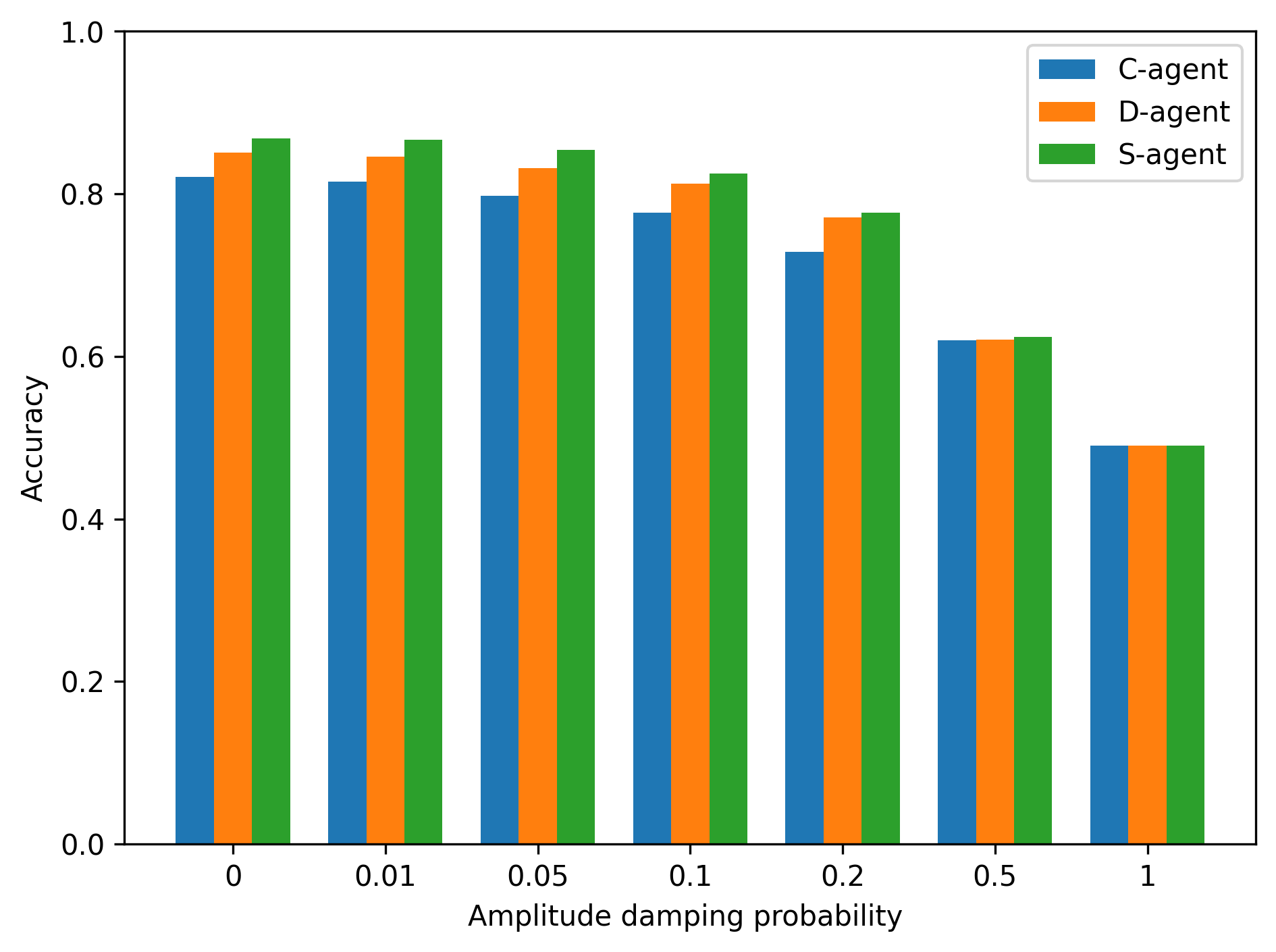}} 
    \vspace{-10pt}
    \caption{Accuracy across different noise models for agents trained on low penalty ($X=0.05$).
}
    \label{fig:noise_0.05}
\end{figure*}

Fig.~\ref{fig:noise_0.05} reports the accuracy of all agents under increasing noise strength when trained in the low-penalty environment ($X=0.05$).

Across all three noise models, bit-flip, depolarizing, and amplitude damping, all agents exhibit high accuracy under low-to-moderate noise levels. As the noise strength increases, a gradual degradation in accuracy is observed for all agents, reflecting the increasing distortion of the measurement statistics.

Both D-agent and S-agent consistently outperform the C-agent across most noise levels, maintaining higher accuracy under identical noise conditions. This indicates that the hybrid policies are able to select measurement strategies that preserve informative statistics even in the presence of channel noise. Consequently, the hybrid agents maintain higher accuracy than the classical baseline across noiseless and moderately noisy conditions. However, at extreme noise strengths, the accuracy of all agents approaches random performance (approximately $0.5$).

\section{Quantum-assisted Authentication}\label{sec:applications}

The strong performance of the proposed approach, characterized by high accuracy and low resource consumption, demonstrates its potential for practical applications. In particular, the ability to reliably extract information using only two quantum copies demonstrates resource efficiency. This advantage is most pronounced for the S-agent, which consistently achieves robust performance with minimal quantum resources. In the following, we demonstrate how our proposed QRL environment can be used for quantum-assisted authentication. 

Since the proposed environment is built around a quantum challenge–response mechanism, a natural application is quantum-assisted authentication, where the protocol serves as an auxiliary component within a two-factor authentication system. In this setting, quantum states are used to encode authentication challenges, while a trained RL agent, whether classical or quantum, interacts with the received states to infer the hidden response. In our implementation, we recommend the use of the S-agent, which offers a balance between accuracy and resource efficiency. By leveraging quantum phenomena, the proposed approach can enhance the security of authentication protocols beyond purely classical methods. 
We consider a challenge–response authentication setting in which Alice acts as the challenger and Bob as the verifier. The protocol proceeds as follows:

\begin{itemize}
    \item \text{Phase 1: Enrollment (Offline)}. Alice and Bob agree on a shared angular domain $\mathcal{D}$. As described in Section~\ref{sec:QCRC}, Alice encodes the hidden bit using a phase selected from a $\pi$-length interval. Accordingly, Alice and Bob must agree on this angular range during enrollment. While the interval length is fixed, there exist infinitely many possible choices for the domain $\mathcal{D}$, which remains secret between the communicating parties. Bob then trains a RL agent to infer the hidden bit $b$ using only two quantum copies per challenge. This training setup mirrors the operating regime achieved by the S-agent, which is capable of reliably extracting the hidden bit using as few as two qubits.

    \item \text{Phase 2: Classical authentication}. Alice and Bob first perform a classical authentication step using conventional methods such as a password or a PIN.
    \item \text{Phase 3: Quantum Challenge}. Upon successful classical authentication, Alice prepares a quantum challenge and encodes the hidden bit $b$ into a quantum state using an angle selected from the agreed angular domain $\mathcal{D}$. For each challenge, Alice prepares $N=2$ identical qubits encoding the same hidden bit. Finally, the qubits are transmitted to Bob over a quantum channel.

    \item \text{Phase 4: Verification}. Bob applies his trained agent to interact with the received quantum states and infer the hidden bit $b'$. Authentication is accepted if the inferred bit matches the original challenge, i.e., $b' = b$, with an acceptance threshold of $0.8$ to account for noise encountered during quantum state transmission.

\end{itemize}

For an adversary Eve to successfully infer the hidden bit, she must intercept the quantum states transmitted to Bob over the quantum channel. In particular, Eve would need access to the two identical qubits prepared for each challenge. However, duplicating these quantum states is prohibited by the no-cloning theorem, which fundamentally limits Eve’s ability to obtain sufficient copies without detection. In addition, Eve must know the secret angular domain $\mathcal{D}$ agreed upon during the offline enrollment phase in order to train an appropriate inference agent. If Eve instead employs an agent trained on a different angular domain, her inference performance degrades significantly and falls below random guessing. In our experiments, such a mismatch results in an accuracy of approximately $0.178$, indicating that Eve is effectively misled and unable to extract meaningful information from the intercepted quantum states.

\section{Conclusion} \label{sec:conclusion}
In this work, we introduced a quantum RL environment formulated as a challenge-response task with hidden information. The proposed environment is intrinsically quantum in nature, as the agent’s observations and interactions arise from measurements performed on quantum states whose parameters encode the information to be inferred. By explicitly incorporating resource constraints in the form of limited quantum state copies and interaction penalties, the environment provides a controlled testbed for studying RL behavior under realistic quantum limitations.

We investigated three agent realizations: a purely classical agent (C-agent), a hybrid quantum-classical actor-critic agent (D-agent), and a simple lightweight hybrid agent (S-agent) that combines shallow quantum feature extraction with a classical decision network. Through extensive experiments, we analyzed the trade-off between inference accuracy and quantum resource consumption under both high and low-penalty scenarios. The results demonstrate that while all agents can achieve high accuracy when resources are abundant, meaningful performance differences emerge under strict interaction constraints.

Specifically, the lightweight S-agent consistently achieved reliable inference using as few as two quantum state copies, outperforming both the classical baseline and the deeper quantum policy. This finding highlights that increased quantum circuit depth does not necessarily translate to improved performance and that compact hybrid policies can offer a favorable balance between expressivity, stability, and resource efficiency. Robustness evaluations under realistic quantum noise models further showed that the hybrid agents maintain improved inference performance compared to the C-agent across a broad range of noise strengths.

Beyond benchmarking agent performance, the proposed environment offers a flexible foundation for exploring security-oriented applications. We discussed its relevance to quantum-assisted authentication, where controlled access to quantum states and classical side information enables implicit access control mechanisms. 

Another potential application for our proposed environment is \emph{information hiding}. In such a setting, classical bits are encoded into quantum states using secret phase angles. The resulting quantum states are stored in quantum memory, while the corresponding angular domain is stored separately as classical side information. Information extraction is only possible for an authorized party that has access to both the stored quantum states and the associated angular domain. This knowledge enables the extractor to train an RL agent tailored to the encoding parameters, which can then interact with the quantum states to reliably recover the hidden bits. Without simultaneous access to both the quantum states and the correct angular domain, any attempt to extract the hidden information yields incorrect results, effectively concealing the embedded data. This dual-dependence on quantum memory and classical side information highlights the potential of RL-based quantum challenge–response mechanisms as a viable tool for information hiding.


Overall, this work positions quantum challenge-response environments as a promising direction for studying QRL, providing insight into how hybrid learning agents can effectively operate under practical quantum constraints. In the future, similar penalty-based interaction mechanisms could be explored in other NISQ-era applications to explicitly regulate quantum resource consumption. In such settings, step penalties may be used to limit the number of circuit executions or measurement shots. Further extensions may include adaptive adversarial settings, the investigation of formal security guarantees, and experimental implementations.

\bibliographystyle{IEEEtran}
\bibliography{ref.bib}
\end{document}